\documentclass[reprint,amsmath,amssymb,aps,superscriptaddress,floatfix,longbibliography,prb]{revtex4-2}
\usepackage{graphicx}
\usepackage{dcolumn}
\usepackage{bm}
\usepackage{color}
\usepackage[caption=false]{subfig}
\usepackage{booktabs}

\newcommand*\LCC{Li$_2$CoCl$_4$}
\newcommand*\CCl{CoCl$_{2}\cdot$2H$_2$O}

\begin{document}

\title{Applied-field magnetic structure and spectroscopy shifts of the effective spin-$\frac{1}{2}$, $XY$-like magnet \LCC}

\author{Zachary W. Riedel}
\affiliation{Department of Materials Science and Engineering, University of Illinois Urbana-Champaign, Urbana, IL 61801, USA}
\affiliation{Materials Research Laboratory, University of Illinois Urbana-Champaign, Urbana, IL 61801, USA}

\author{Mykhaylo Ozerov}
\email{ozerov@magnet.fsu.edu}
\affiliation{National High Magnetic Field Laboratory, Florida State University, Tallahassee, FL 32310, USA}

\author{Stuart Calder}
\affiliation{Neutron Scattering Division, Oak Ridge National Laboratory, Oak Ridge, TN 37831, USA}

\author{Daniel P. Shoemaker}
\email{dpshoema@illinois.edu}
\affiliation{Department of Materials Science and Engineering, University of Illinois Urbana-Champaign, Urbana, IL 61801, USA}
\affiliation{Materials Research Laboratory, University of Illinois Urbana-Champaign, Urbana, IL 61801, USA}

\begin{abstract}
Insulators containing chains of magnetic transition metal cations provide platforms for probing spin-$\frac{1}{2}$ dynamics and quantum critical behavior. \LCC\ contains edge-sharing CoCl$_6$ octahedra that form chains along the crystallographic $c$ axis and orders antiferromagnetically at zero field, but questions remain about its applied-field magnetic structure and the Co$^{2+}$ spin state. Here, we show with neutron diffraction on a polycrystalline sample how the anti-aligned chains of cobalt moments begin to transition to a ferromagnetic state above 1.6~T. Further, using magnetic resonance absorption measurements and noninteracting spin models, we reveal the strongly anisotropic nature of the Co$^{2+}$ ion's $XY$-like magnetic behavior ($g_{\parallel}=2.77$ and $g_{\perp}=5.23$) and its $J=\frac{1}{2}$ ground state. We, therefore, supply the magnetic structures and anisotropic description needed to explore the dynamics of the field-driven magnetic phases, laying the foundation for further experimental and theoretical studies.
\end{abstract}

\maketitle

\section{Introduction}
Low-dimensional magnetic materials frequently provide access to quantum behavior, and scattering studies on unexplored, spin-$\frac{1}{2}$ materials are critical for validating low-dimensional magnetism models that may be extended to higher-dimensional systems \cite{vasiliev2018milestones}. Of note here, several compounds with chains of edge-sharing Co$^{2+}$$X_6$ polyhedra have exhibited quantum criticality \cite{coldea2010quantum,larsen2017spin,wang2018quantum}. We identified \LCC\ as a low-dimensional magnetism candidate with a dimensional analysis toolkit \cite{karigerasi2018uncovering}. \LCC\ contains Co$^{2+}$ octahedrally coordinated to Cl$^{-}$, forming edge-sharing polyhedral chains along the crystallographic $c$ axis. 

The ionic conductivity of Li$_2$$M$Cl$_4$ compounds has been studied extensively \cite{kanno1981ionic,kanno1983phase,schmidt1984fast,kanno1987new,kanno1988structure,kanno1988structureLi2FeCl4,kanno1988synthesis,lutz1988ionic,wussow1989lattice,lutz1995schnelle}, but magnetic property measurements have been limited to susceptibility data used to determine the spin states of Mn$^{2+}$ and Fe$^{2+}$ in cubic Li$_2$MnCl$_4$ \cite{kanno1984ionic} and Li$_2$FeCl$_4$ \cite{cerisier1986conducteurs}. For orthorhombic \LCC, high temperature susceptibility data indicate \mbox{spin-$\frac{3}{2}$} (high-spin) behavior with significant unquenched orbital angular momentum, and heat capacity data point to a ground-state Kramers doublet ($J=\frac{1}{2}$) dominating low-temperature behavior \cite{riedel2023zero}. Other high-spin Co$^{2+}$ chlorides exhibit similar effective \mbox{spin-$\frac{1}{2}$} behavior, leading to fruitful studies of spin excitations and quantum criticality \cite{achiwa1969linear,kjems1975spin,montfrooij2001spin,larsen2017spin,mena2020thermal}. Therefore, we were encouraged to map the magnetic phases of \LCC.

At zero field, \LCC\ orders antiferromagnetically with $P_{C}bam$ magnetic space group symmetry. The Co$^{2+}$ magnetic moments align within edge-sharing CoCl$_6$ chains and anti-align between nearest neighbor chains. The magnetic susceptibility and heat capacity shift with increasing field, leading to three proposed magnetic phase regions separated by transitions near 1.6~T and 3.5~T at 2~K \cite{riedel2023zero}. The magnetic structures of the applied-field regions are unknown. The low-field region is antiferromagnetic, likely similar to the zero-field magnetic ordering. The intermediate-field region may be ferrimagnetic, with chain rotations reminiscent of the structurally similar compound \CCl\ \cite{cox1966neutron,weitzel1974neutron}, though the data were also consistent with a spin-flop reorientation. Finally, in the high-field region, not sharply delineated from the intermediate-field one, the \LCC\ moment chains are likely aligned, but with no well-defined Curie temperature in the magnetic susceptibility. 

Here, we use polycrystalline \LCC\ to probe the applied-field regions' magnetic structures with neutron diffraction, particularly that of the previously ambiguous intermediate-field region. Additionally, we extract several parameters from applied-field infrared spectroscopy data, such as the spectroscopic $g$-factors, to discuss the role of spin-orbit coupling and octahedral distortion on the magnetic properties of \LCC.

\section{Materials and methods}
We prepared polycrystalline \LCC\ as previously described \cite{riedel2023zero}. We ground LiCl (99.9\%, Alfa Aesar) and CoCl$_2$ (99.7\%, Alfa Aesar) under argon, sealed the mixture under vacuum, heated it to 550$^{\circ}$C at 10$^{\circ}$C/min, held for 12~h, then cooled to room temperature at 10$^{\circ}$C/min. We ensured that the highly hygroscopic LiCl and \LCC\ powders remained under inert atmospheres during synthesis and characterization.

We collected neutron diffraction data on the HB-2A beamline at the High Flux Isotope Reactor at Oak Ridge National Laboratory (ORNL). To avoid grain reorientation under applied field, we pressed pellets of \LCC\ for the diffraction experiment. Under argon, we pressed 1.5~g of powder into around 30 1/8'' diameter pellets. We then sealed the pellets under vacuum, transported them to ORNL, and loaded them under helium into a 6~mm diameter vanadium canister. The top of the canister was packed with aluminum foil before being hermetically sealed. 

We collected constant wavelength (2.41~\AA) neutron diffraction data using a Ge(113) monochromator with the instrument in an open-open-12' pre-mono, pre-sample, pre-detector collimation. The sample was cooled to base temperature (1.5--1.6~K) under zero field before collecting powder scans that covered 2$\theta$ angles from 6.025$^{\circ}$ to 127.525$^{\circ}$ in 0.05$^{\circ}$ steps with ${\mu}_{0}H$~=~0, 0.8, 1.5, 2.5, 3.5, 4.5, and 5.5~T. We also collected a zero-field scan at 20~K (above the ordering temperature). The cryomagnet contributed a significant reflection at 2$\theta$~=~19.9$^{\circ}$. Sweeping scans monitored select magnetic reflections with increasing field or temperature, also after zero-field cooling. To analyze the neutron diffraction data, we used \textsc{GSAS-II} \cite{toby2013gsas} in combination with the Bilbao Crystallographic Server's \textsc{k-Subgroupsmag} program \cite{perez2015symmetry}, generating magnetic structure images with \textsc{VESTA} \cite{momma2011vesta}.

We performed magneto-infrared measurements at the infrared (IR) spectroscopy facility at the National High Magnetic Field Laboratory with a 17.5~T vertical-bore superconducting magnet coupled to a Bruker Vertex 80v FTIR spectrometer. In an argon glovebox, a 3--5~mg polycrystalline sample was bonded with n-eicosane and loaded between two polypropylene layers to protect it from oxygen and moisture. We placed the sample in a Voigt geometry so that the incoming IR light was perpendicular to the magnetic field. A composite Si bolometer (IR Labs) collected the transmitted IR light with a spectral range of 10--1100~cm$^{-1}$ (0.3--33~THz) and a resolution of 0.3~cm$^{-1}$ (9~GHz). Additionally, we placed a low-pass THz filter (QMC Instruments) with a 120~cm$^{-1}$ cutoff frequency in front of the sample to increase sensitivity in the far-IR (FIR) range and to minimize radiative heating. IR data were collected near 5~K. 

The FIR transmission spectra were divided by a reference spectrum constructed from the maximum transmission at each wavenumber across all spectra measured at different magnetic fields. Such normalized transmission spectra, $T_B$, are only sensitive to intensity changes induced by the magnetic field and eliminate contributions from magnetic-field-independent vibrational absorption and instrument artifacts. The magnetic resonance absorption spectrum is then calculated as $A=-log(T_B)$, and results were analyzed using the EPR analysis package \textsc{EasySpin} \cite{stoll2006easyspin}.

\section{Results and Discussion}
\subsection{Zero- and low-field magnetic structures}\label{sec:LF}
We first confirmed that the zero-field magnetic structure matches that determined previously. The data refines well to the antiferromagnetic structure with $P_{C}bam$ (BNS no. 55.363) magnetic space group symmetry (see supplemental material \cite{supplement}). For the magnetic space group $P_{C}bam$, only reflections where $h+k\neq2n$ are allowed, whereas the nuclear $Cmmm$ space group only allows reflections where $h+k=2n$, making it simple to identify the magnetic ordering reflections. The magnetic structure is commensurate, with the propagation vector \textbf{k}~=~(1,0,0). The magnetic moments are aligned within the Co$^{2+}$ chains and anti-aligned between nearest neighboring chains along the nuclear cell's $ab$-plane diagonal. The antiferromagnetic interchain interactions lead to the previously reported cusp in the magnetic susceptibility and the negative Curie-Weiss temperature \cite{riedel2023zero}. The refined moment is 2.10(3)~$\mu_{\mathrm{B}}$. Though the moment in this study is slightly lower than the previously refined value 2.19(4)~$\mu_{\mathrm{B}}$, the refinement standard errors are likely underestimated. Since we were able to produce a closer fit to peak intensities in this work, especially nuclear intensity, we will use 2.10~$\mu_{\mathrm{B}}$ here. The 20~K, zero-field data in the paramagnetic region is included in the supplemental material for comparison \cite{supplement}.

We noticed several low-intensity impurity peaks at 2${\theta}=$~17.0, 30.4, 46.6, 53.0, and 68.5$^{\circ}$ ($Q=$~0.77, 1.37, 2.06, 2.32, and 2.93~\AA$^{-1}$), the largest impurity peak being only 0.8\% the intensity of the largest \LCC\ peak. The $Q=$~0.77~\AA$^{-1}$ peak also appears in X-ray diffraction data after exposing a sample to air for 15--30~s before sealing it under vacuum \cite{supplement}. The reflections do not change with temperature, indicating they are nonmagnetic, and they do not influence data analysis.

\begin{figure}[h]
    \centering
    \includegraphics[width=0.85\columnwidth]{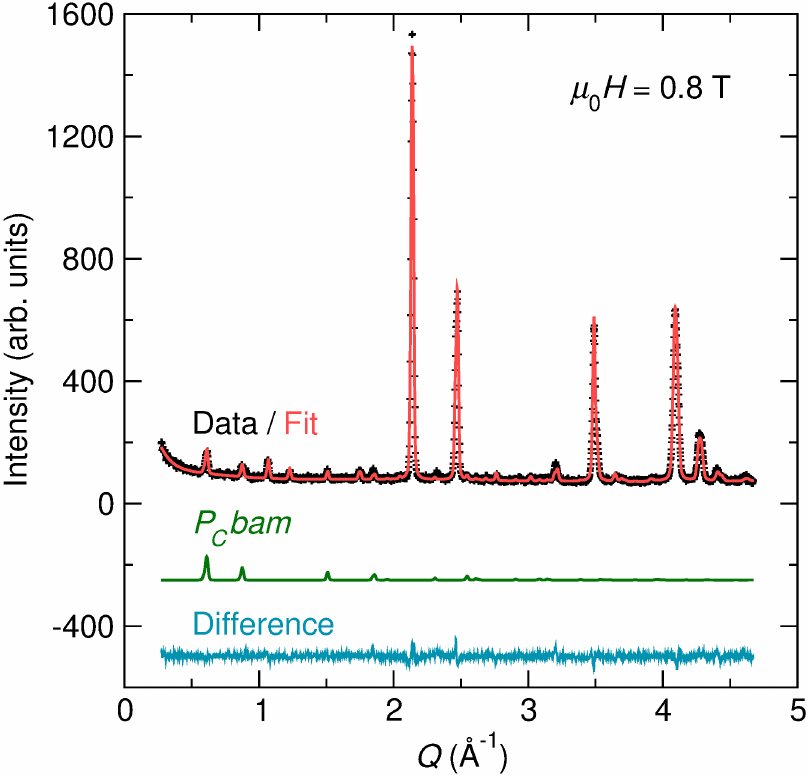}
    \caption{The neutron powder diffraction data collected at 1.5~K and 0.8~T refine well to the antiferromagnetic magnetic phase also observed at zero field with a 2.02(3)~$\mu_{\mathrm{B}}$ moment along the \textit{c} axis for Co$^{2+}$. The refined lattice parameters are $a=$7.1711(3), $b=$10.2393(5), and $c=$3.6111(2)~\AA.}
    \label{fig:0p8T_fit}
\end{figure}

Prior magnetic susceptibility and heat capacity data indicate that the low-field magnetic structure ($<$~1.6~T) matches the zero-field structure. At 0.8~T, we observe the same magnetic reflections as in the zero-field case, associated with a loss of $C$-centering, but with slightly decreased intensity. Refinements give an unconstrained magnetic moment of 2.02(3)~$\mu_{\mathrm{B}}$ (Fig.~\ref{fig:0p8T_fit}), indicating some magnetic intensity may become paramagnetic background, though restricting the moment to 2.10~$\mu_{\mathrm{B}}$ gives a nearly identical fit. With the field increased to 1.5~T, just below the 1.6~T transition to the intermediate-field phase, we can again refine the data well to the $P_{C}bam$ structure with a constrained 2.10~$\mu_{\mathrm{B}}$ moment or a refined moment of 1.96(3)~$\mu_{\mathrm{B}}$. See the supplemental material \cite{supplement} for the 1.5~T refinement. 

\subsection{Field-induced moment rotation}\label{sec:SF}
Between the 1.5~T and 2.5~T scans, the $h+k\neq2n$ antiferromagnetic structure reflections, e.g. (010), (100), and (120), lose significant intensity while the $h+k=2n$ reflections allowed by the $C$-centered nuclear cell, e.g. (110) and (020), gain intensity. The intensity shifts indicate a spin reorientation to a \textbf{k}~=~\textbf{0} magnetic structure. 
Because of the nuclear cell's $C$ centering, only ferromagnetic arrangements are consistent with \textbf{k}~=~\textbf{0} ordering.
Unlike for \CCl\ \cite{cox1966neutron,weitzel1974neutron}, we do not observe new peaks associated with a ferrimagnetic phase where one-third of the chains are anti-aligned. The difference in behavior may be due to the distance between anti-aligned chains in the zero-field structures (5.6~\AA\ for \CCl\ and 6.2~\AA\ for \LCC), producing a weaker next-nearest-neighbor (antiferromagnetic) exchange interaction in \LCC, and thus a lower barrier to overcoming the interaction with applied field. Instead, in \LCC\ a spin reorientation begins to align the chains of Co$^{2+}$ magnetic moments. The exact nature of the moment reorientation cannot be determined. If the reorientation involves a spin-flop, as previously proposed, a magnetic field applied along the $c$ axis would lead to a temporary ``flop" to a hard axis perpendicular to $c$ before a strong enough field produced a ferromagnetic arrangement of moments along $c$. Since the magnetic field is not applied along a specific crystallographic direction, we cannot confirm nor repudiate the presence of a spin-flop phase.

The randomly oriented structural grains lead to a slow evolution of the magnetic diffraction intensity with increasing field and a mixture of magnetic phases.
For some of the randomly oriented structural grains in the sample, the magnetic field will align with the crystallographic axis preferred by the magnetic anisotropy, leading to the lowest-energy ferromagnetic arrangement and coherent magnetic scattering intensity. In contrast, for grains where the magnetic field is not aligned with the anisotropy-preferred direction, the competing perturbations will lead to moment canting off the anisotropy-preferred axis toward the field, explaining the previously observed net magnetization for a polycrystalline sample \cite{riedel2023zero}. The magnetic scattering reflects the random distribution of these grains with canted moments and thus their randomly distributed moment orientations.

Several ferromagnetic models can successfully capture the additional magnetic intensity at high fields. The simplest model uses the maximal magnetic space group $Cm'm'm$ (BNS no. 65.485), which constrains the cobalt magnetic moments along the crystallographic $c$ axis. Additional maximal magnetic groups are considered in the supplemental material \cite{supplement}. Fig.~\ref{fig:unit_cells} shows the $P_{C}bam$ and $Cm'm'm$ magnetic unit cells along with the nuclear cell. Considering lower-symmetry, commensurate magnetic structures with canted moments does not improve fits to the ferromagnetic intensity. For example, the $C$2$'/m'$ space group symmetry (BNS no. 12.62) is the highest symmetry \textbf{k}~=~\textbf{0} magnetic space group possible for \LCC\ that allows for moment components along $a$ and $c$. Refinements to the 5.5~T data using $C$2$'/m'$ ($M_{\mathrm{a}}$~=~1.5(2)~$\mu_{\mathrm{B}}$, $M_{\mathrm{c}}$~=~1.7(1)~$\mu_{\mathrm{B}}$, $|M|=2.3(1)~\mu_{\mathrm{B}}$) gave a similar fit to the simpler $Cm'm'm$ model \cite{supplement}. Likewise, an arrangement with $P\bar{1}$ magnetic symmetry (BNS no. 2.4) allows for moment components along all three crystallographic directions and gives a similar fit ($|M|=2.3(2)  ~\mu_{\mathrm{B}}$). While there is ambiguity in the refinements regarding the $Cm'm'm$, $C2'/m'$, and $P\bar{1}$ magnetic symmetries, the $P\bar{1}$ scenario is likely more consistent with the polycrystalline sample's evolution due to the previously discussed randomness of the structural grains' orientations relative to the applied magnetic field, leading to an average magnetic structure with moment components along all three crystallographic directions.

The best refinement models above 1.6~T involve both the antiferromagnetic $P_{C}bam$ magnetic phase and a ferromagnetic \textbf{k}~=~\textbf{0} structure.
Using the simplest model ($Cm'm'm$ symmetry) allows us to track the evolution of the magnetic structure of \LCC\ at magnetic fields greater than 1.6~T by refining the data with contributions from the $P_{C}bam$ antiferromagnetic structure and the $Cm'm'm$ ferromagnetic structure. The moment is constrained to be equal in the two magnetic phases when both are refined. The full $Q$ range refinements for the 2.5, 3.5, 4.5, and 5.5~T data are shown in Fig.~S4 \cite{supplement}. Table~\ref{tab:refinements} contains the refined phase fractions of the $P_{C}bam$ and $Cm'm'm$ magnetic structures and the refined cobalt magnetic moment for each magnetic field.

\begin{figure*}
    \centering
    \subfloat{\includegraphics[width=0.66\columnwidth]{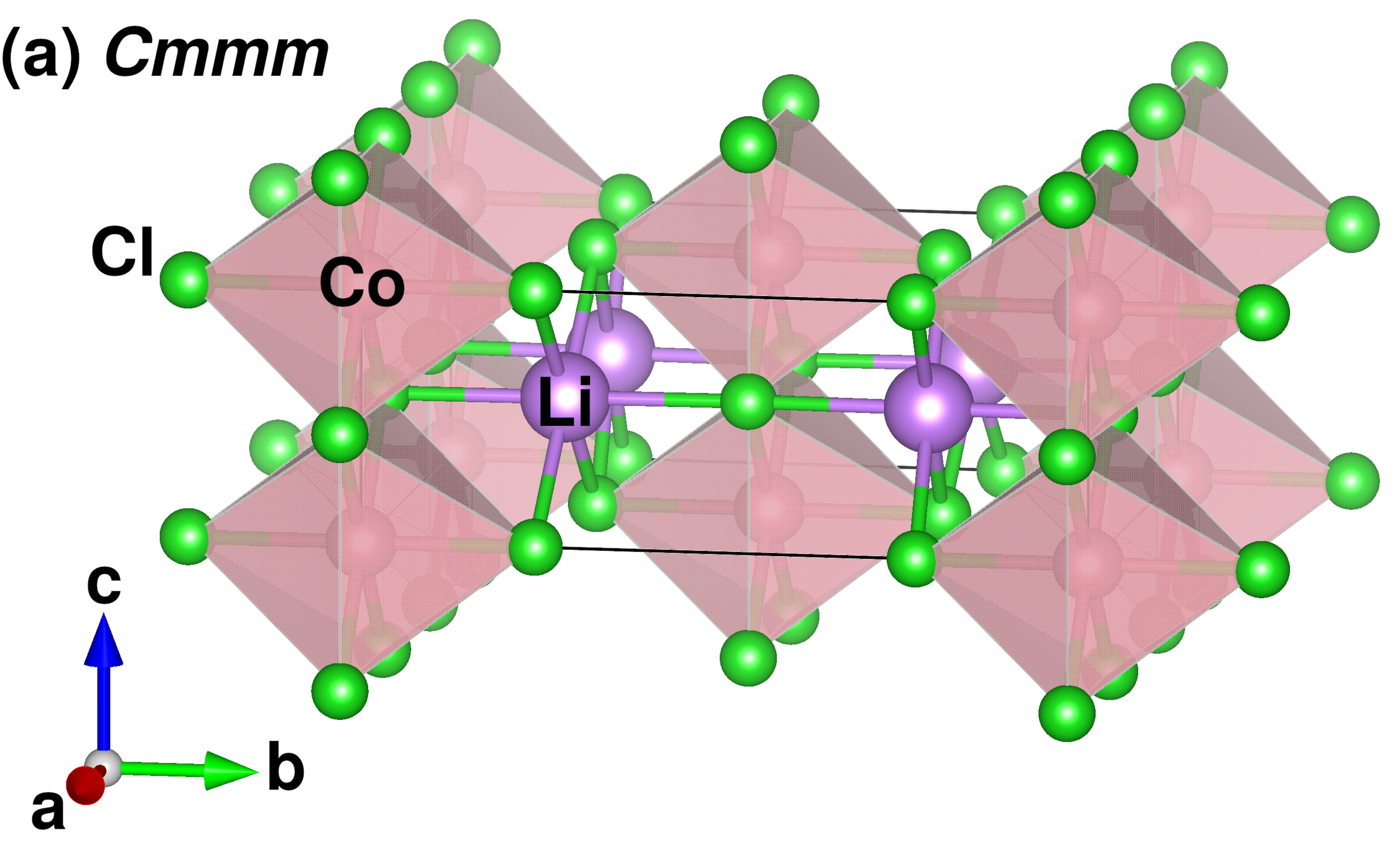}}
    \hfill
    \subfloat{\includegraphics[width=1.33\columnwidth]{figures/mag_cells_flatten_1.pdf}}
    \caption{\small The nuclear and magnetic cells of \LCC\ are identified by their space groups. (a) The nuclear (and paramagnetic) unit cell contains chains of edge-sharing Co--Cl octahedra along $c$ connected by chains of Li--Cl octahedra along $a$. (b) At zero field and below $\sim$7~K, Co$^{2+}$ chains are ferromagnetic with antiferromagnetic interchain interactions, breaking the cell's $C$-centering. For applied fields above 1.5--1.6~T, ferromagnetic ordering emerges as the antiferromagnetic interactions are overcome.}
    \label{fig:unit_cells}
\end{figure*}

\begin{figure}
    \centering
    \includegraphics[width=0.85\columnwidth]{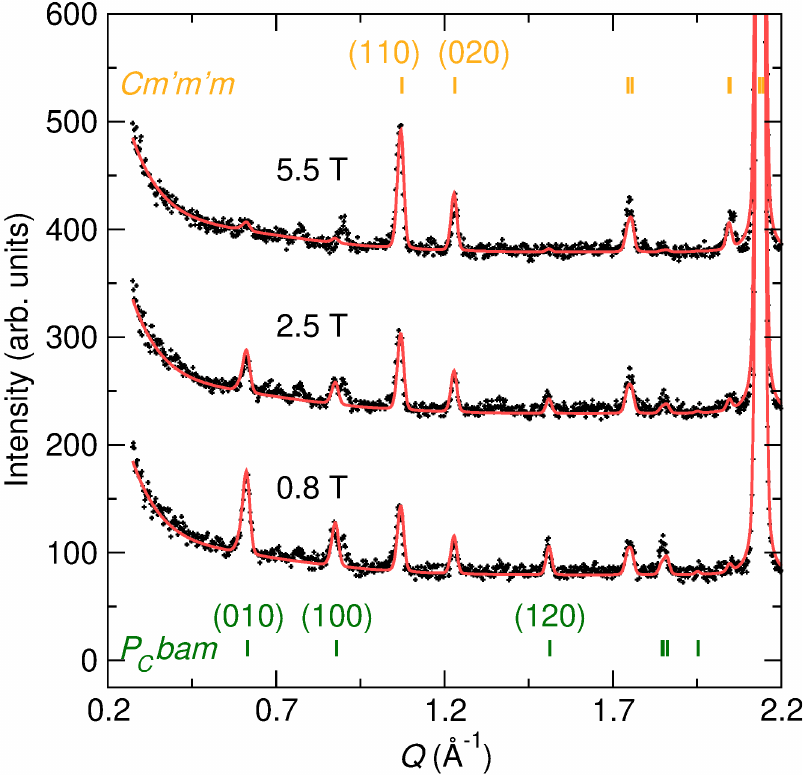}
    \caption{Neutron powder diffraction data from the three regions (antiferromagnetic, spin-flop, ferromagnetic) are shown with several reflections labeled to highlight the phase contribution changes with increasing field. All unfit peaks are from the instrument/impurity features mentioned in the main text.}
    \label{fig:0p8T_2p5T_5p5T_fits}
\end{figure}

To highlight the changing magnetic contributions with increasing field, Fig.~\ref{fig:0p8T_2p5T_5p5T_fits} compares the 0.8, 2.5, and 5.5~T refinements, corresponding to the previously described low-, intermediate-, and high-field regions. Since the intermediate-field region involves the greatest competition between the antiferromagnetic and ferromagnetic interactions, the refined magnetic moment is significantly reduced for the 2.5 and 3.5~T data. Likewise, while most of the refinements are qualitatively identical if the cobalt magnetic moment is constrained to the zero-field value of 2.10~$\mu_{\mathrm{B}}$, the 2.5 and 3.5~T refinements significantly change (Fig.~S5 \cite{supplement}). At higher fields where the ferromagnetic phase dominates, the  refined moment returns, within error, to its zero-field value, and constraining the moment to 2.10~$\mu_{\mathrm{B}}$ does not influence the fit.

\begin{table}[t]
\small
\centering 
\caption{\label{tab:refinements} Rietveld refinement results for base temperature, constant applied field scans are listed. Scans below 1.6~T only include the antiferromagnetic $P_{C}bam$ phase, while those above 1.6~T also include the ferromagnetic $Cm'm'm$ phase. The Co$^{2+}$ magnetic moment is along the $c$ axis in both cases.} 
\begin{tabular}{c c c}
\midrule \parbox{1cm}{Field \\(T)} & \parbox{4cm}{Magnetic Phase \\(\% $P_{C}bam$ - \% $Cm'm'm$)} & \parbox{2.5cm}{Refined Moment \\(${\mu}_{\mathrm{B}}$)} \\
\midrule
0 & 100~-~0 & 2.10(3) \\
0.8 & 100~-~0 & 2.02(3) \\
1.5 & 100~-~0 & 1.96(3) \\
2.5 & 81(6)~-~19(6) & 1.59(7) \\
3.5 & 42(3)~-~58(3) & 1.76(6) \\
4.5 & 17(2)~-~83(2) & 2.07(6) \\
5.5 & 10(2)~-~90(2) & 2.05(6) \\
\end{tabular}
\end{table}

\begin{figure}
    \centering
    \includegraphics[width=0.85\columnwidth]{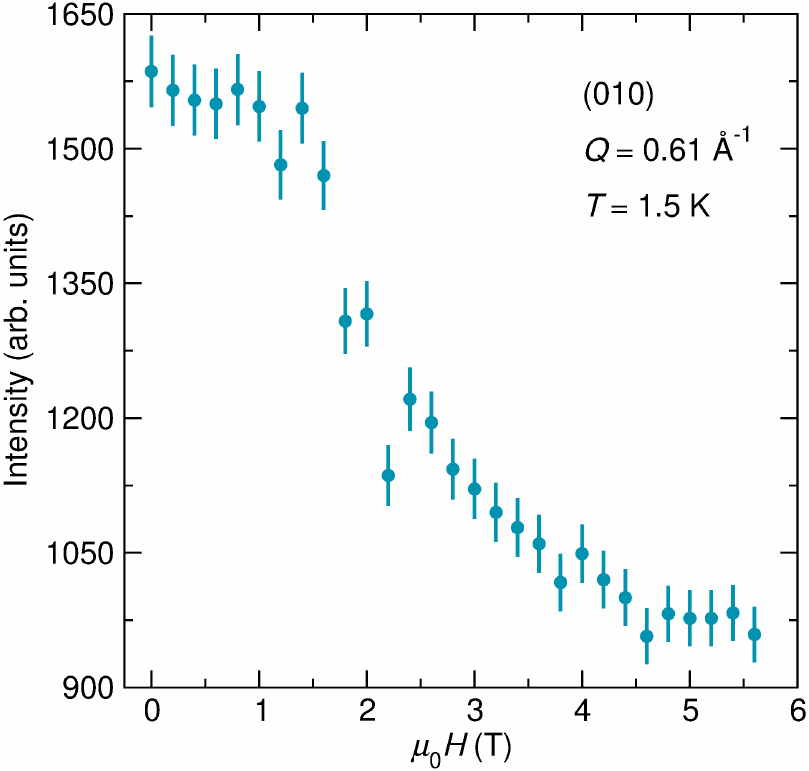}
    \caption{With increasing field, the antiferromagnetic (010) reflection intensity is constant before decreasing with the emergence of the ferromagnetic phase near 1.5--1.6~T. Above 4.5~T, the intensity levels off as the ferromagnetic phase dominates.}
    \label{fig:010_field_sweep}
\end{figure}

The emergence of the ferromagnetic phase with increasing field can be monitored more closely by tracking the relatively high intensity (010) reflection associated with the antiferromagnetic phase (Fig.~\ref{fig:010_field_sweep}). Above 1.6~T, the (010) intensity drops until the intensity levels off near 4.5~T. The drop in (010) reflection intensity at 1.6~T matches the slope change of the magnetization curve, the susceptibility's leveling off below the critical temperature, and the emergence of a hump in the magnetic heat capacity alongside a sharper peak \cite{riedel2023zero}. The suppression of the heat capacity peak along with the emergence of a hump, therefore, is due to the reduction of antiferromagnetic regions and the short-range aligning of chains in newly formed ferromagnetic regions of the randomly oriented polycrystalline sample. While tracking the (010) reflection, we also captured changes in lower intensity reflections with the other diffractometer detectors: the antiferromagnetic (100) reflection and the nuclear/ferromagnetic (131) reflection. Both show the same trend of increasing ferromagnetic intensity and decreasing antiferromagnetic intensity above 1.5--1.6~T \cite{supplement}.

By the time the field reaches 5.5~T, the (010) reflection intensity has approached the experimental background, the majority of the sample is no longer antiferromagnetic, and the magnetic anisotropy likely favors an arrangement of moments along the $c$ axis. Using the previous DFT results for a 3.5~K experimental cell, the $Cm'm'm$ moment arrangement is 0.62~meV/atom higher in energy than the $P_{C}bam$ moment arrangement \cite{riedel2023zero}, so, trivially, for an isolated 2.10~$\mu_{\mathrm{B}}$ moment along $c$ paired with an anti-aligned 5.5~T field, the energy gain for flipping the moment is 1.34~meV (\(2 {\mu_{\mathrm{Co}}} B\)). The proposed magnetic structure change is, therefore, energetically reasonable. In the ferromagnetic-dominated field region, we tracked the (110) transition with increasing temperature (Fig.~S8). Up to 20~K, we do not observe a sharp drop or leveling off of the intensity but rather a nearly linear decrease in intensity. Similarly, the magnetic susceptibility at high field, even up to 7~T (Fig.~S8), does not show the typical sharp rise of a paramagnetic to ferromagnetic transition, suggesting that the high-field magnetic ordering is driven by field overcoming the material's interchain antiferromagnetic exchange interactions instead of by dominant interchain ferromagnetic exchange interactions.

\subsection{Spin model from high-field IR spectroscopy}\label{sec:IR}
\begin{figure*}
\centering
    \includegraphics[width=1.9\columnwidth]{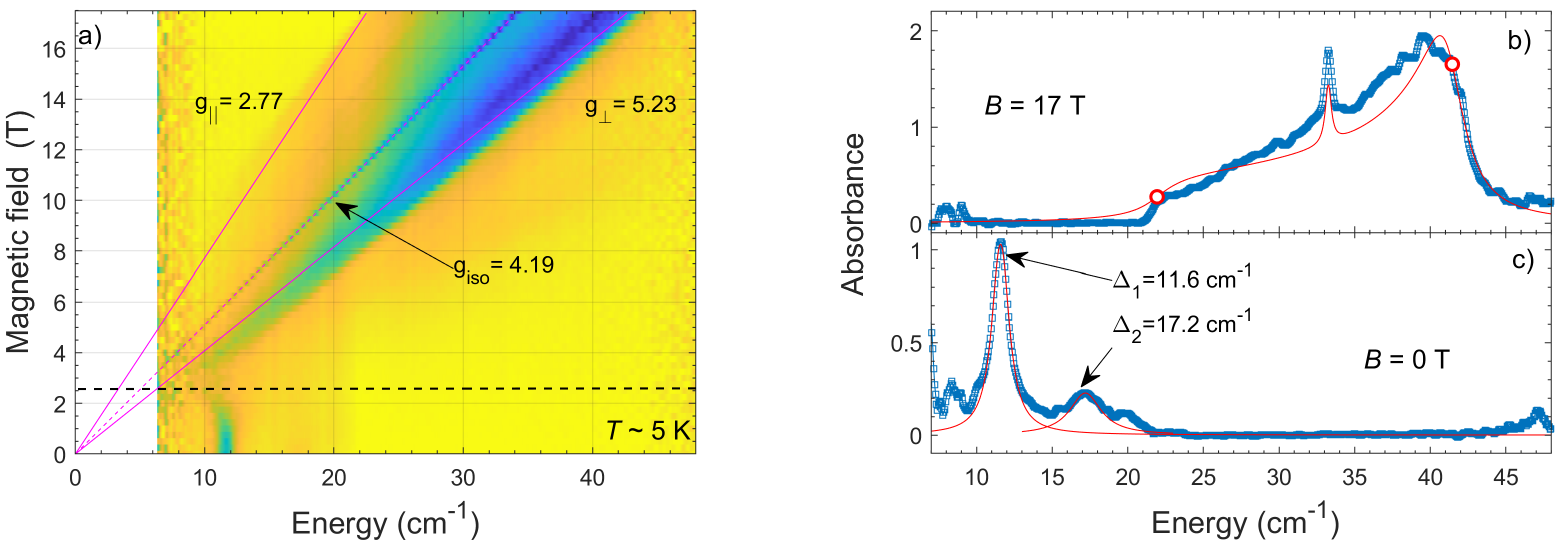}
    \caption{(a) The magnetic resonance absorption heat map at 5~K is shown with magenta lines representing the resonance conditions for different $g$-factors for a noninteracting spin-$\frac{1}{2}$ model. The black dashed line corresponds to the transition to the high-field magnetic phase from magnetic susceptibility data in Ref.~\cite{riedel2023zero} ($\sim$2.5~T at 5~K). (b) A heat map slice at 17~T (blue dots) shows agreement with a simulated powder spectrum (red line) with 98\% of the intensity from a spin-$\frac{1}{2}$ site with an anisotropic $g$-factor ($g_{\perp}$ and $g_{\parallel}$, magenta points) and 2\% from a site with an isotropic $g$-factor of 4.19. (c) Lorentzian peak fits (red curves) to a heat map slice at zero field demonstrate the nearly two-fold linewidth difference in the spin-gap peaks.
    }
    \label{fig:FIRMS_5K}
\end{figure*}

To explore the spin dynamics of \LCC\ under high applied fields, we collected IR spectroscopy data up to 17.5~T. At lower energies within the detection range, the normalized magnetic resonance absorption contains a broad continuum (Fig.~\ref{fig:FIRMS_5K}a). The broad continuum reveals the strongly anisotropic nature of the magnetic moment of the Co$^{2+}$ cation in the \LCC\ powder sample. At magnetic fields significantly above the moment saturation field, the Zeeman energy dominates the antiferromagnetic spin-spin interactions, and spin dynamics converge toward those of free-ion spin precession \cite{nagamiya1955antiferromagnetism}. 
For many low-dimensional cobalt compounds with octahedral ligand environments, the magnetic properties are well-described by an effective spin-$\frac{1}{2}$ model due to a well-separated ground state Kramers doublet \cite{oguchi1965theory,shiba2003exchange,Wang2020,Yang2023}. Therefore, as a first approximation, we use a noninteracting spin-$\frac{1}{2}$ framework to extract $g$-factors from the high-field IR data.

\subsubsection{Noninteracting spin-$\frac{1}{2}$ model}
In the free spin-$\frac{1}{2}$ framework, an applied magnetic field splits the $m={+\frac{1}{2}}$ and $m={-\frac{1}{2}}$ states (Zeeman splitting) by the energy $E=g\mu_\mathrm{B}B$, where $E$ can be directly observed as a peak in the absorption spectrum. The parameter $g$ is the effective $g$-factor, whose value depends on the orientation of the magnetic field relative to the anisotropy axis. Due to the randomly oriented grains in the polycrystalline sample, the absorption associated with the Zeeman splitting is distributed over a frequency range that broadens with magnetic field, resulting in the continuum in Fig.~\ref{fig:FIRMS_5K}a. The asymmetric lineshape of the absorption spectrum (Fig.~\ref{fig:FIRMS_5K}b) indicates a system with axial symmetry, and indeed the CoCl$_6$ octahedra in \LCC\ are slightly compressed along the $b$ axis [ratio of the Co-Cl bond length along $b$ to the length within the $ac$ plane of 0.986(2)], producing a tetragonal distortion of the octahedra. The continuum is bounded by resonance conditions $g_\mathrm{\parallel}\mu_\mathrm{B}B$ and $g_\mathrm{\perp}\mu_\mathrm{B}B$ with more intensity on the $g_\mathrm{\perp}$ side \cite{weil2007epr}, where $\parallel$ and $\perp$ are relative to the compressed CoCl$_6$ axis. Since the continuum edges at the resonance energies associated with $g_\mathrm{\parallel}$ and $g_\mathrm{\perp}$ are smeared by peak broadening, we simulated the absorption spectrum using a powder-averaged lineshape and varied the $g$-factors to obtain a match (Fig.~\ref{fig:FIRMS_5K}b). The fit anisotropic $g$-factors are $g_{\parallel}$~=~2.77 and $g_{\perp}$~=~5.23. The corresponding resonance lines confine the magnetic absorption spectrum for a free cobalt ion, but the experimental data diverge from the lines below $\sim$10~T, where exchange interactions produce a more complex picture of collective spin excitations (Fig.~\ref{fig:FIRMS_5K}a).

Also of note is the narrow feature within the broader continuum of Fig.~\ref{fig:FIRMS_5K}a at high fields. The lineshape is well-fit by a single peak, indicating an isotropic $g$-factor of 4.19. The peak may originate from a Co$^{2+}$ site with minimal axial distortion. A similar peak was observed in the X-band EPR spectrum of LiCoO$_2$ and was ascribed to a high-spin Co$^{2+}$ site on the surface of polycrystalline sample \cite{hu2021multifunctional,hu2023MRL}. Powder spectrum simulations that consider a mixture of isotropic (surface) and anisotropic (bulk) Co$^{2+}$ sites (Fig.~\ref{fig:FIRMS_5K}b) indicate that 2\% of the Co$^{2+}$ sites are isotropic. We will regard the fraction as a small paramagnetic impurity and will ignore it while interpreting other magnetic property measurements of powder samples. 
The fit $g$-factors in the noninteracting spin-$\frac{1}{2}$ model are significantly larger than the isotropic, free-electron value of $g=2$. The discrepancy can be well explained by an effective Hamiltonian that accounts for spin-orbit coupling.

\subsubsection{Effective spin-orbit coupling model}
To build a spin-orbit coupling model, we first consider the ligand environment and paramagnetic susceptibility of Co$^{2+}$ in \LCC. The Co$^{2+}$ cations are octahedrally coordinated to Cl$^{-}$ anions, and the magnetic susceptibility indicates that the Co$^{2+}$ cations are high-spin with unquenched orbital angular momentum \cite{riedel2023zero}, i.e. significantly influenced by spin-orbit coupling (SOC).
The anisotropic magnetism of a high-spin Co$^{2+}$ cation in a local octahedral environment is well-studied \cite{abragam1951theory,lines1963magnetic,mabbs1973magnetism,kahn1993molecular,shiba2003exchange}. The ground state is a $^{4}T_{\mathrm{1} g}$($^4$F) orbital triplet with $L_{\mathrm{eff}}$~=~1 and $S$~=~$\frac{3}{2}$, and the following Hamiltonian describes the low-energy diagram \cite{kahn1993molecular}:
\begin{equation} \label{eq:Hamiltonian}
    H = -{\lambda}\mathbf{L}{\cdot}{\alpha}{\cdot}\mathbf{S}-\Delta(L^{2}_{\mathrm{z}}-\frac{2}{3})+{\mu_\mathrm{B}}\mathbf{B}(g_{\mathrm{e}}\mathbf{S}-{\alpha}\mathbf{L})
\end{equation}
$\lambda$ is the spin-orbit coupling constant, $\Delta$ is the crystal field energy due to axial distortion, \textbf{B} is the magnetic field vector, $g_{\mathrm{e}}$ is the free-electron $g$-factor, and $\alpha$ is the orbital reduction factor. The orbital reduction factor quantifies the covalence of metal-ligand bonds and admixture of higher-energy orbital states, e.g. $^{4}T_{\mathrm{1} g}$($^4$P). Without axial distortion or a magnetic field, the first-order spin-orbit coupling splits the twelve-fold degenerate $^{4}T_{\mathrm{1} g}$ state, making a $J$~=~$\frac{1}{2}$ doublet the lowest energy level. The doublet is separated by $9\lambda/4$ from the first excited quartet state (49.7~meV assuming a free-ion spin-orbit coupling constant of 178~cm$^{-1}$)  \cite{abragam1951theory,goodenough1968spin,mabbs1973magnetism}. The well-separated, ground-state doublet, therefore, produces effective spin-$\frac{1}{2}$ behavior at low temperatures with an isotropic $g$-factor of \mbox{$g_0$=(10+2$\alpha$)/3} with $\alpha$ values ranging from 1 (strong crystal field) to 1.5 (weak crystal field), resulting in isotropic $g$-factors between 4 and 4.33 \cite{lloret2008magnetic}. 

Next, axial distortion should be considered because of the aforementioned tetragonal distortion of the CoCl$_6$ octahedra in \LCC.
Axial distortion of the ideal octahedral environment of the Co$^{2+}$ ion further splits the $J$ manifold into six Kramers doublets and leads to anisotropy of the $g$-factor. For distortion values ($\Delta$) small relative to the spin-orbit exchange constant ($\lambda\alpha$), the $g$-factor anisotropy is approximately \cite{lloret2008magnetic}:
\begin{align}
\begin{split}
    g_{\perp} = g_{\mathrm{0}} + \frac{8\Delta}{27{\lambda}{\alpha}}(\alpha + 2)  \\    g_{\parallel} = g_{\mathrm{0}} - 2\frac{8\Delta}{27{\lambda}{\alpha}}(\alpha + 2)
    \end{split}
\end{align}
The average $g$-factor in this case remains independent of the distortion effects since $g_{\mathrm{avg}}=(2g_{\perp}+g_{\parallel})/3=g_0$. Using the fit spectroscopic $g$-factors $g_{\parallel}=2.77$ and $g_{\perp}=5.23$, we can calculate a $g_{\mathrm{avg}}$ of 4.41, which agrees well with the predicted isotropic $g$-factor of 4.33 for high-spin Co$^{2+}$ ion in the weak crystal field limit. The calculated $g_\mathrm{avg}$ also corroborates the neutron diffraction moment, which we would expect to equal $g_{\mathrm{avg}}J~\mu_{\mathrm{B}} = 2.2~\mu_{\mathrm{B}}$ if $J=\frac{1}{2}$. Therefore, the observed $g$-factors for \LCC\ confirm a $J=\frac{1}{2}$ (Kramers doublet) ground state.

\begin{figure*}
    \centering
    \includegraphics[width=0.9\textwidth]{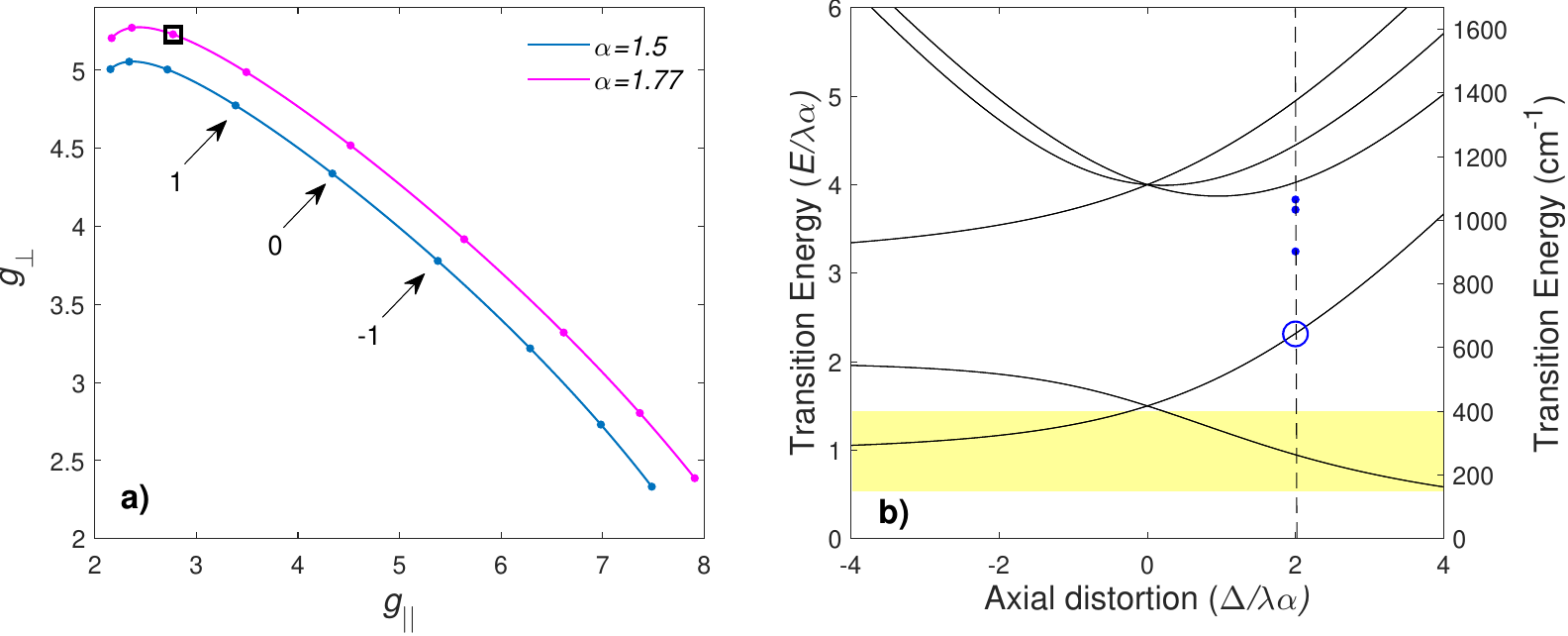}
    \caption{(a) Curves show the change in $g$-factor anisotropy with axial distortion for orbital reduction factors of 1.5 and 1.77 with points indicating integer values of $\Delta/(\lambda\alpha)$. (b) The spectrum for a free Co$^{2+}$ ion is calculated for $\alpha$~=~1.77 as a function of E/(${\lambda}{\alpha}$) (left-hand axis) or assuming ${\lambda}$~=~157~cm$^{-1}$ (right-hand axis). The dashed line matches the distortion derived from the experimental $g$-factor anisotropy. The shaded area is opaque to FIR radiation, preventing observation of transitions to the first excited Kramers doublet.
    }
    \label{fig:distortion_SOC}
\end{figure*}

Having confirmed the applicability of the spin-orbit coupling model, we can solve the Hamiltonian (Eq.~\ref{eq:Hamiltonian}) to further probe the interplay between the spin-orbit exchange constant and crystal field axial distortion in \LCC. Using the \textsc{EasySpin} package \cite{stoll2006easyspin}, we tuned the ratio of the axial distortion to the spin-orbit exchange constant, $\Delta$/($\lambda\alpha$), and calculated the resulting changes in $g_\mathrm{\parallel}$ and $g_\mathrm{\perp}$ (Fig.~\ref{fig:distortion_SOC}a). The experimental $g$-factor values are well-described by a $\Delta/(\lambda\alpha)$ ratio of 2 and a reduction factor ($\alpha$) of 1.77, which is close to the weak crystal field case of 1.5. The relatively small $\Delta/(\lambda\alpha)$ ratio \cite{abragam1951theory} is indicative of a system where spin-orbit coupling plays a significant role \cite{kahn1993molecular}. The reduction factor is reasonably similar to that of other cobalt salts \cite{abragam1951theory}, though the value is outside the expected limits of [1,1.5] for a noninteracting system, highlighting a deficiency of neglecting the magnetic coupling between neighboring Co$^{2+}$ ions \cite{lloret2008magnetic}.

\subsubsection{Observed excitations predicted by the SOC model}
\begin{figure}
    \centering
    \includegraphics[width=0.9\columnwidth]{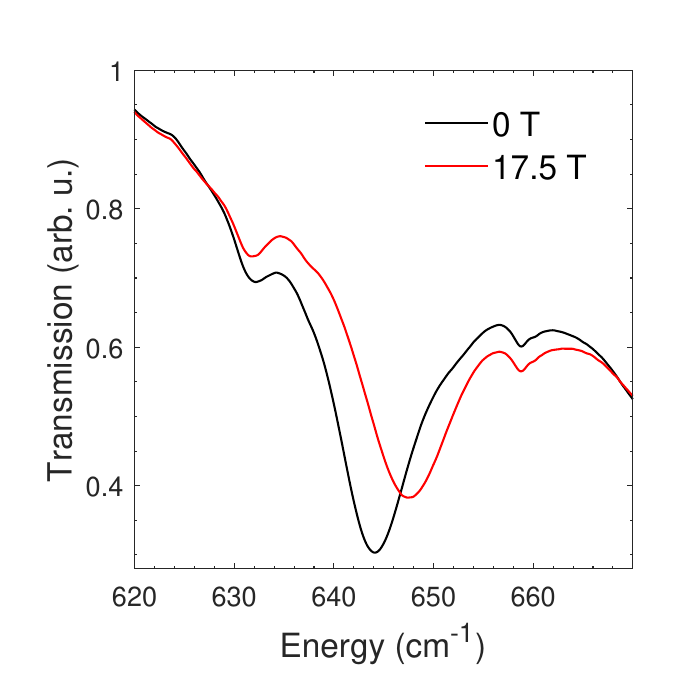}
    \caption{The transmission spectra of the \LCC\ powder sample measured at 0 and 17.5~T. The peak at 644~cm$^{-1}$ shifts under high magnetic fields, while the small peaks, likely associated with phonons, remain at the same position. 
    }
    \label{fig:FIRMS_high_energy}
\end{figure}

To assess the SOC model at higher energies in the absorption spectrum, we calculated the energy difference between the ground-state and higher-energy Kramers doublets as a function of the $\Delta/(\lambda\alpha)$ ratio (Fig.~\ref{fig:distortion_SOC}b) to compare with the experimental data. Since an applied magnetic field further splits the Kramers doublets, we can distinguish spectroscopic peaks associated with the transitions from other spectroscopic features in the experimental spectrum. For example, a pronounced peak at 644~cm$^{-1}$ shifts and broadens under an applied magnetic field (Fig.~\ref{fig:FIRMS_high_energy}). Assuming $\Delta/(\lambda\alpha)=2$, we attribute this energy to a transition from the ground state to the second excited Kramers doublet, denoted by a blue circle in Fig.~\ref{fig:distortion_SOC}b. Using the previously estimated value of $\alpha$ (1.77), this energy yields a spin-orbit coupling constant of $\lambda$~=~157~cm$^{-1}$, consistent with the 178~cm$^{-1}$ value for a free Co$^{2+}$ ion \cite{abragam1951theory}.
The transition to the first excited state, however, is not observed in the spectrum because of a strong absorption band between 200 and 400~cm$^{-1}$. Nevertheless, additional field-dependent spectral features (blue dots in Fig.~\ref{fig:distortion_SOC}b) were observed at higher energies (Fig.~S9 \cite{supplement}) and are likely associated with the remaining inter-Kramers doublet transitions. The model, therefore, qualitatively captures the anisotropic physics of Co$^{2+}$ in \LCC\ at higher energies in the measured absorption spectrum.

\subsubsection{Paramagnetic magnetization and magnetic susceptibility}
\begin{figure*}
    \centering
    \includegraphics[width=0.9\textwidth]{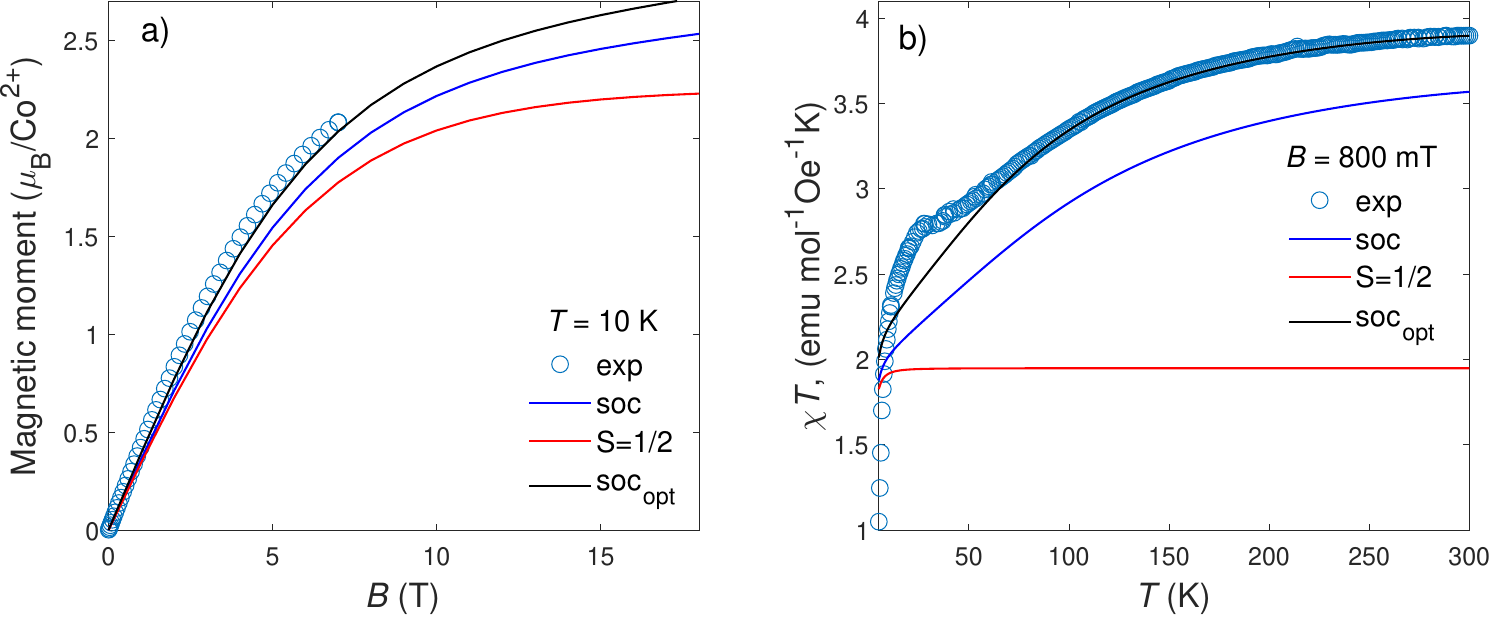}
    \caption{Experimental (a) magnetization data at 10~K and (b) susceptibility data at 0.8~T on \LCC\ powder (``exp") are presented for a spin-$\frac{1}{2}$ model using $g_{\perp}$~=~5.23 and $g_{\parallel}$~=~2.77 (``S=1/2"); a spin-orbit coupling model using the spectroscopically determined parameters $\alpha$~=~1.77, $\Delta/(\lambda\alpha)$~=~2, and $\lambda$~=~157~cm$^{-1}$ (``soc"); and a spin-orbit coupling model using the parameters $\alpha$~=~2, $\Delta/(\lambda\alpha)$~=~2.2, and $\lambda$~=~110~cm$^{-1}$, adjusted to match the experimental data (``soc$_\mathrm{opt}$"). 
    }
    \label{fig:modeling_mag}
\end{figure*}

In addition to predicting IR excitations, the SOC model can be compared to experimental magnetization and magnetic susceptibility data, where the contributions from low-lying excited states are relevant. Fig.~\ref{fig:modeling_mag} compares the experimental data in the paramagnetic region to \textsc{EasySpin} calculations utilizing Eq.~\ref{eq:Hamiltonian} for three cases: a noninteracting spin-$\frac{1}{2}$ model with the $g$-factors from our high-field spectroscopy data; a spin-orbit coupling model with the $\alpha$, $\Delta$, and $\lambda$ parameters fit to our high-field spectroscopy data; and a spin-orbit coupling model with spectroscopic parameters optimized to fit the experimental magnetization and magnetic susceptibility. All three models qualitatively match the shape of the magnetization data and indicate a saturated moment near 2-2.75~${\mu}_{\mathrm{B}}$, but they differ significantly in describing the susceptibility data. While both spin-orbit coupling models reasonably match the low- and high-temperature slopes of the susceptibility, the spin-$\frac{1}{2}$ model only captures the low-temperature behavior. Improvements to the spin-orbit coupling model would need to consider higher-order spin-orbit interactions and electronic structure effects.

\subsubsection{Spin gaps below the saturation field}
The approximation of free spin dynamics in high magnetic fields, of course, breaks down at lower magnetic fields where the spin-spin interactions are dominant in determining the excitation spectrum. At zero field, we observe a strong peak at 11.6~cm$^{-1}$ and a weaker satellite at 17.2~cm$^{-1}$ (Fig.~\ref{fig:FIRMS_5K}c) that we attribute to spin gaps in the collective spin excitations of \LCC. Within a free-ion framework, the observed 11.6~cm$^{-1}$ gap could be interpreted as the zero-field energy of a $S=\frac{3}{2}$ system. Exploring this possibility, we found moderate agreement between the experimental data and a $S=\frac{3}{2}$ Hamiltonian with a single-ion anisotropy of $D=5.8$~cm$^{-1}$ and anisotropic $g$-factors of $g_{\perp}=4.50$ and $g_{\parallel}=4.16$ (Fig.~S10~\cite{supplement}). However, to observe a first excited Kramers doublet $\sim$10~cm$^{-1}$ would require extremely high axial distortions (Fig.~\ref{fig:distortion_SOC}b), which would be inconsistent with the model's $g$-factors (Fig.~\ref{fig:distortion_SOC}a). 

The 11.6~cm$^{-1}$ and 17.2~cm$^{-1}$ spin gaps also cannot be explained using the phenomenological theory of antiferromagnetic resonance (AFMR) for collinear magnets with two-axis magnetic anisotropy \cite{glazkov2005magnetic}. The model predicts two excitation modes attributed to spin-flop and spin-flip reorientations with increasing magnetic field. The magnetic field softens the AFMR resonance energies until the spin gaps close at the critical fields, which are directly related to the zero-field energies by the $g$-factor. But for a spin-flop field of 1.6~T ($B_\mathrm{sf}$) and a resonance frequency of 11.6~cm$^{-1}$, the necessary $g$-factor would be $g=\omega/(\mu_\mathrm{B}B_\mathrm{sf})=15.5$, significantly above our observed $g$-factors. For the features to be AFMR modes, the zero-field gap would have to be much smaller or the spin-flop field much larger.
Therefore, understanding the low-field spectroscopic data requires a more detailed microscopic model involving anisotropic spin-spin exchange interactions.

\subsection{Evidence of $XY$-like anisotropy}
A notable implication of our IR spectroscopy data is $XY$-like magnetism of the Co$^{2+}$ cation in \LCC\ since $g_{\perp}>g_{\parallel}$. Transition energies calculated with the SOC model (Fig.~\ref{fig:distortion_SOC}b) predict an energy gap between the two lowest doublets of 264~cm$^{-1}$ (32.7~meV), indicating that Co$^{2+}$ ions' behavior in \LCC\ is dominated by the ground-state Kramers doublet at low temperatures. This state can be mapped to an effective spin-$\frac{1}{2}$ state with an exchange interaction, $JS_1S_2$, between next-neighbor spins exhibiting anisotropy of the type $\beta=J_z/J_x\neq1$. Solving the spin-orbit Hamiltonian (Eq.~\ref{eq:Hamiltonian}) using spectroscopically determined parameters enables an estimation of the magnetic coupling anisotropy, yielding $\beta= 0.41$ for \LCC\ (see the supplemental material for details \cite{supplement} and references \cite{abragam1951theory,oguchi1965theory,achiwa1969linear,shiba2003exchange} therein). This value of $\beta$ classifies \LCC\ as a system with easy-plane anisotropy, consistent with $g_\mathrm{\perp}>g_\mathrm{\parallel}$. Most Co-based chain materials with octahedral Co$X_6$ coordination, such as CsCoCl$_3$ \cite{achiwa1969linear}, CoCl$_2\cdot$2H$_2$O \cite{Hansen2022}, CoNb$_2$O$_6$ \cite{Wang2020}, [Co(NCS)$_2$(4-methoxypyridine)$_2$]$_n$ \cite{rams2020single}, Co(N$_2$H$_5$)$_2$(SO$_4$)$_2$ \cite{calder2022magnetic}, and (Ba,Sr)Co$_2$V$_2$O$_8$ \cite{Yang2023}, exhibit Ising anisotropy ($\beta>1$). The tetragonal distortion of the CoCl$_6$ octahedron in \LCC, in contrast to the trigonal distortion in those systems, may serve as the underlying mechanism for $XY$ anisotropy.

The neutron diffraction data are also compatible with, though not definitive regarding, easy-($ac$-)plane anisotropy. 
The data suggest a preference for moment alignment along the $c$ axis, but there were no observed reflections that violate a rotation through the $a$ axis during the spin reorientation, and the refinement with moment components along $a$ and $c$ ($C$2$'/m'$ space group) was nearly identical to that with moments only along $c$. 
In contrast, the zero-field structure has no moment component along $b$, and under an applied field, the maximal magnetic space group with moments solely along $b$ was incompatible with the magnetic intensity. These features of the neutron data suggest that the $c$ axis or $ac$ plane may be the material's easy axis/plane, which is consistent with the IR data. If the system has easy-axis anisotropy along $c$ instead of easy-plane anisotropy, this is still compatible with the $XY$-anisotropy description, which neglects spin-spin exchange interactions that would, presumably, distinguish the $a$ and $c$ axes. The SOC model was also overparameterized if we introduced additional anisotropy to try to distinguish $a$ and $c$, leaving $XY$ anisotropy as the best description with the available data.

Qualitatively, the magnetic heat capacity of \LCC\ at temperatures just above the long-range magnetic ordering \cite{riedel2023zero} displays a hump similar to that of the predicted heat capacity for an isolated spin-$\frac{1}{2}$, $XY$ chain with $J_x=J_y$, $J_z=0$, and an applied field along the $z$ direction \cite{katsura1962statistical}. Fitting the heat capacity of Cs$_2$CoCl$_4$ \cite{algra1976heat}, a compound with tetrahedrally-coordinated Co$^{2+}$ which also has a long-range ordering lambda peak and a hump in its magnetic heat capacity, to this model gives an intrachain exchange interaction parameter. However, we could not similarly fit the heat capacity of polycrystalline \LCC, which has appreciable interchain interactions. The hump is, therefore, better understood as stemming from short-range order due to the spatially distributed spin reorientation with increasing field.

\section{Conclusions}
Overall, \LCC\ exhibits strongly anisotropic magnetic properties. At zero field, neutron diffraction shows that the moments within chains of octahedrally coordinated Co$^{2+}$ align ferromagnetically while aligning antiferromagnetically between nearest-neighboring chains. At nonzero fields below 1.6~T, neutron diffraction shows that the antiferromagnetic structure is unchanged, but with increasing magnetic field, the anti-aligned chains undergo a spin reorientation to a ferromagnetic state, likely with a preference for moment alignment along the $c$ axis. Based on refinements including the antiferromagnetic $P_{C}bam$ phase and a $Cm'm'm$ ferromagnetic phase, a majority of the sample becomes ferromagnetic near 3.5~T. The chains do not transition to a ferrimagnetic state at intermediate fields as in \CCl.

The high-field FIR data are well-described by a spin-orbit coupling model that confirms a well-separated Kramers doublet ground state ($J$~=~$\frac{1}{2}$) for the Co$^{2+}$ ion that is approximately 32.7~meV below the first excited state. The $J$~=~$\frac{1}{2}$ ground state and strong spin-orbit coupling explain the qualitative shape of magnetization and susceptibility data in the paramagnetic region, as well as the refined magnetic moment from neutron diffraction, the entropy recovery extracted from previous heat capacity data, and the difficulty in fitting the paramagnetic susceptibility to a purely Curie-Weiss shape. Still, the two spin gaps at 11.6 and 17.2~cm$^{-1}$ in the zero-field spectrum highlight the need for a model that also accounts for the anisotropy of spin-spin interactions. The tetragonal distortion of the CoCl$_6$ octahedra may also promote $XY$-like magnetic interactions, with an estimated interaction parameter ratio $J_z$/$J_x$ of 0.41. $XY$ anisotropy distinguishes \LCC\ from other Co$^{2+}$ chain magnets that have trigonally distorted octahedra and display Ising-like anisotropy.

Further work should explore this anisotropy and its influence on the unexplained spin gaps below the moment saturation field by developing a model of spin-spin interactions between the cobalt ions. The models will be aided by single crystal neutron scattering experiments. Single crystal magnetization and neutron data along the three crystallographic axes would also improve the description of the magnetic behavior, allowing for a more precise determination of the moment saturation field and a definitive description of the spin reorientation at intermediate magnetic fields. \LCC, therefore, offers a rare opportunity to study the combined effects of effective spin-$\frac{1}{2}$ behavior and $XY$-like magnetic anisotropy on spin dynamics in an octahedrally coordinated Co$^{2+}$ system. 

\begin{acknowledgments}
The authors acknowledge the use of facilities and instrumentation supported by NSF through the University of Illinois Materials Research Science and Engineering Center DMR-1720633. A portion of this research used resources at the High Flux Isotope Reactor, a DOE Office of Science User Facility operated by the Oak Ridge National Laboratory. The beam time was allocated to HB-2A (POWDER) on proposal number IPTS-30219. A portion of this work was performed at the National High Magnetic Field Laboratory, which is supported by National Science Foundation Cooperative Agreement No. DMR-2128556 and the State of Florida.
\end{acknowledgments}

\bibliography{Li2CoCl4.bib}

\end{document}


\begin{center}
\Large 
\textbf{Applied-field magnetic structure and spectroscopy shifts of the effective spin-$\frac{1}{2}$, $XY$-like magnet \LCC}\\
\vspace{1em}
Supplemental Material\\
\vspace{1em}
\normalsize
Zachary W. Riedel, Mykhaylo Ozerov, Stuart Calder, and Daniel P. Shoemaker
\end{center}

\vspace{2em}

\section{Additional diffraction data and refinements}
\begin{figure}[h]
    \centering
    \includegraphics[width=0.65\columnwidth]{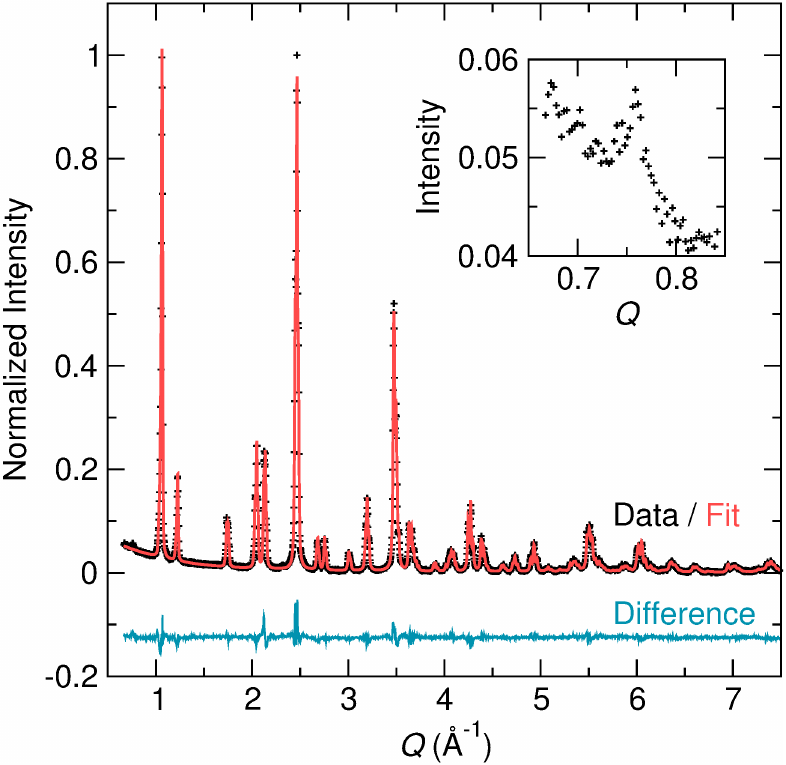}
    \caption{We exposed a portion of the sample used for the neutron powder diffraction experiments to air for 15--30~s before sealing it under vacuum and collecting powder X-ray diffraction data with a Mo-K$\alpha$ source. The data refines well to the known structure, but a low intensity reflection at 0.77~\AA$^{-1}$ appears (see inset). Though this peak also appears in the neutron powder diffraction data, where care was taken to avoid any air exposure, its intensity does not change and is low relative to the reflections from the nuclear and magnetic cells of \LCC.}
    \label{fig:xrd_air}
\end{figure}

\begin{figure}
    \centering
    \includegraphics[width=0.65\columnwidth]{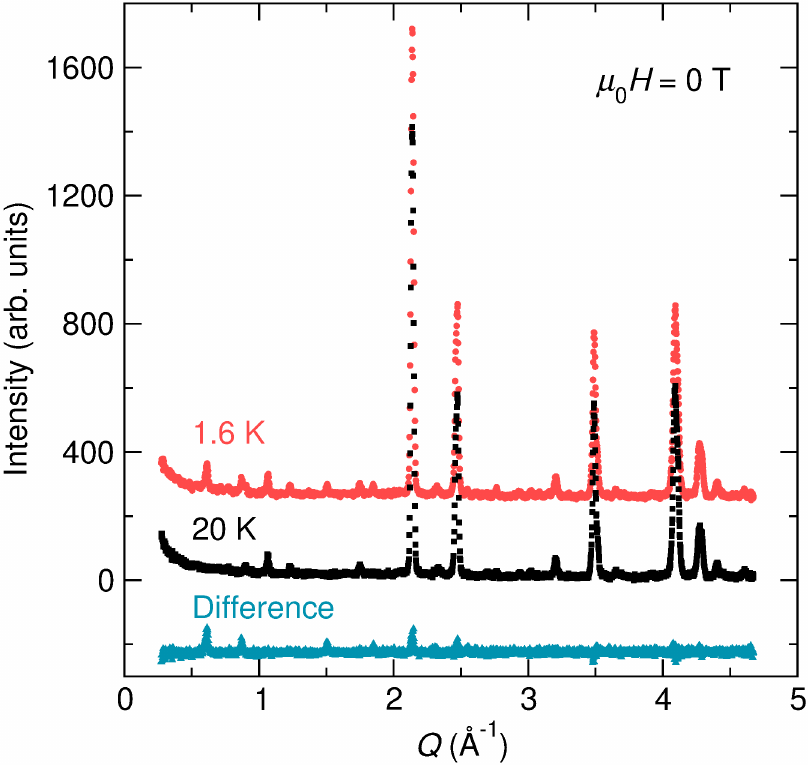}
    \caption{Zero-field neutron powder diffraction data collected at 20~K and 1.6~K are compared, showing the emergence of antiferromagnetic ordering that breaks the nuclear cell's \textit{C}-centering, allowing reflections where $h+k\neq2n$.}
    \label{fig:npd_20K_0T_v_base_0T}
\end{figure}

\begin{table}
\small
\centering 
\caption{\label{tab:fixedmoment} Rietveld refinement results for base temperature, constant magnetic field scans with a constrained moment of 2.10~${\mu}_{\mathrm{B}}$ are listed. $R_\mathrm{w}$ values are included to compare the constrained moment fits to the unconstrained/refined moment fits tabulated in the main text. Scans below 1.6~T only include the antiferromagnetic $P_{C}bam$ phase, while those above 1.6~T also include the ferromagnetic $Cm'm'm$ phase.} 
\begin{tabular}{c c c c}
\midrule \parbox{1cm}{\centering Field \\(T)} & \parbox{4.5cm}{\centering Magnetic Phase \\(\% $P_{C}bam$ - \% $Cm'm'm$)} & \parbox{1cm}{\centering $R_\mathrm{w}$ \\(\%)} & \parbox{4cm}{\centering Refined Moment $R_\mathrm{w}$ \\(\%)}\\
\midrule
0 & 100~-~0 & 5.995 & 5.995 \\
0.8 & 100~-~0 & 5.836 & 5.825 \\
1.5 & 100~-~0 & 5.964 & 5.931 \\
2.5 & 55(2)~-~45(2) & 6.062 & 5.971 \\
3.5 & 35(2)~-~65(2) & 6.007 & 5.964 \\
4.5 & 17(2)~-~83(2) & 5.722 & 5.721 \\
5.5 & 11(2)~-~89(2) & 5.703 & 5.702 \\
\end{tabular}
\end{table}

\begin{figure}
    \centering
    \includegraphics[width=0.65\columnwidth]{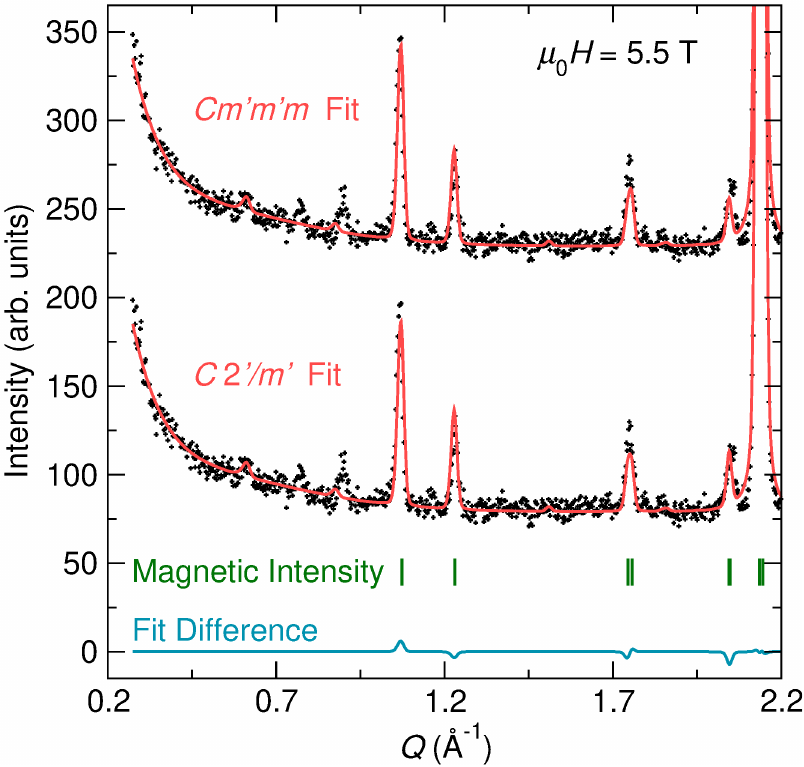}
    \caption{At 5.5~T, the diffraction data can be fit with a small fraction of the antiferromagnetic phase plus a majority ferromagnetic phase with either $Cm'm'm$ symmetry (moment component along $c$) or $C\mathrm{2}'/m'$ symmetry (moment components along $a$ and $c$). The difference between the fits ($Cm'm'm$ fit $-$ $C\mathrm{2}'/m'$ fit) shows minor changes. Both space groups have the same reflection conditions as the nuclear cell; the peak positions are marked with ticks. Unfit peaks are background/impurity peaks discussed in the main text.}
    \label{fig:npd_high_field_options}
\end{figure}

\begin{figure*}
    \centering
    \subfloat{\includegraphics[width=0.4\columnwidth]{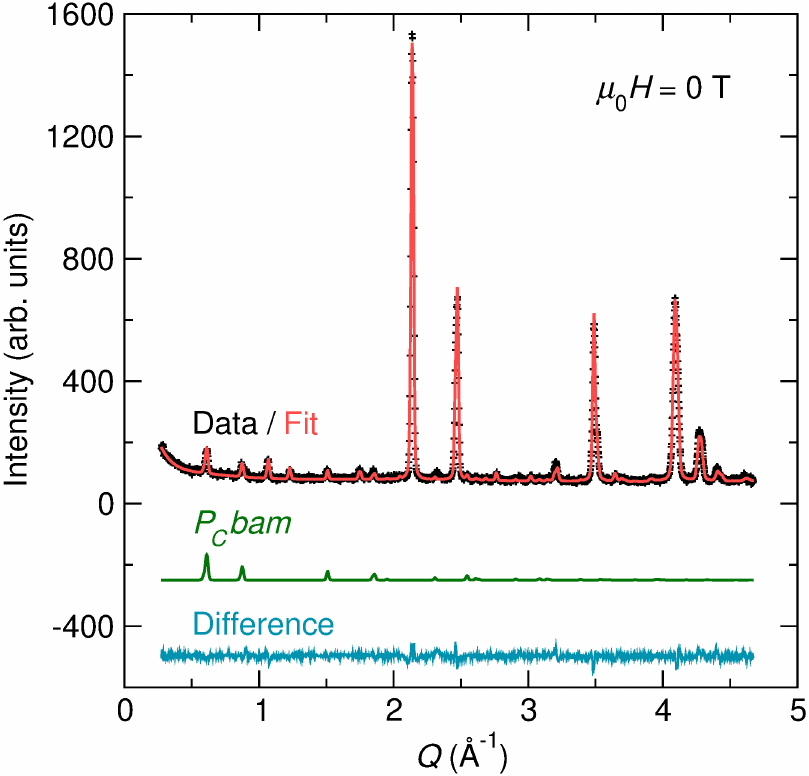}}
    \hspace{1.25cm}
    \subfloat{\includegraphics[width=0.4\columnwidth]{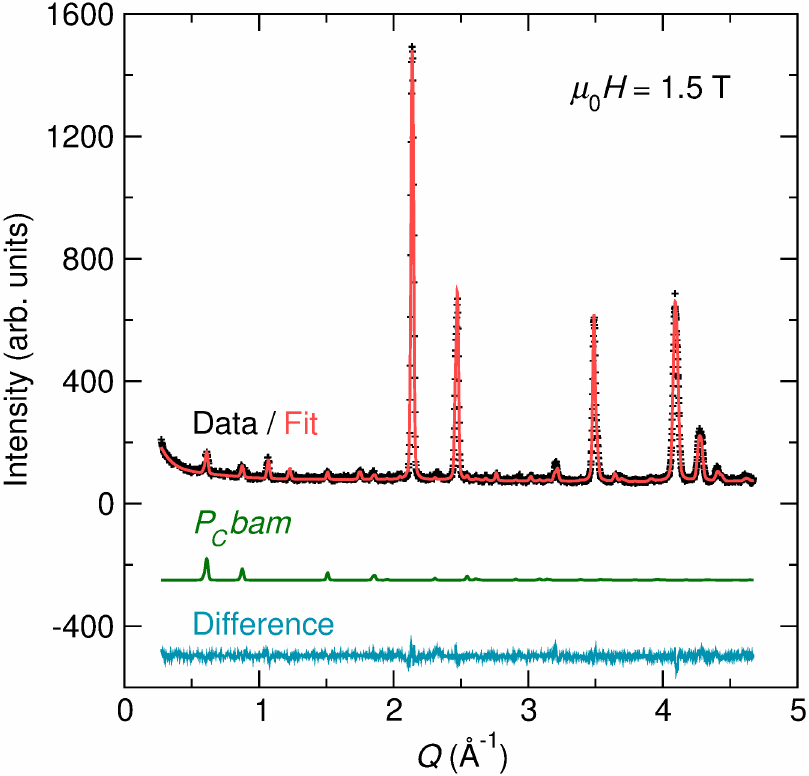}}
    \vspace{0.5cm}
    \subfloat{\includegraphics[width=0.4\columnwidth]{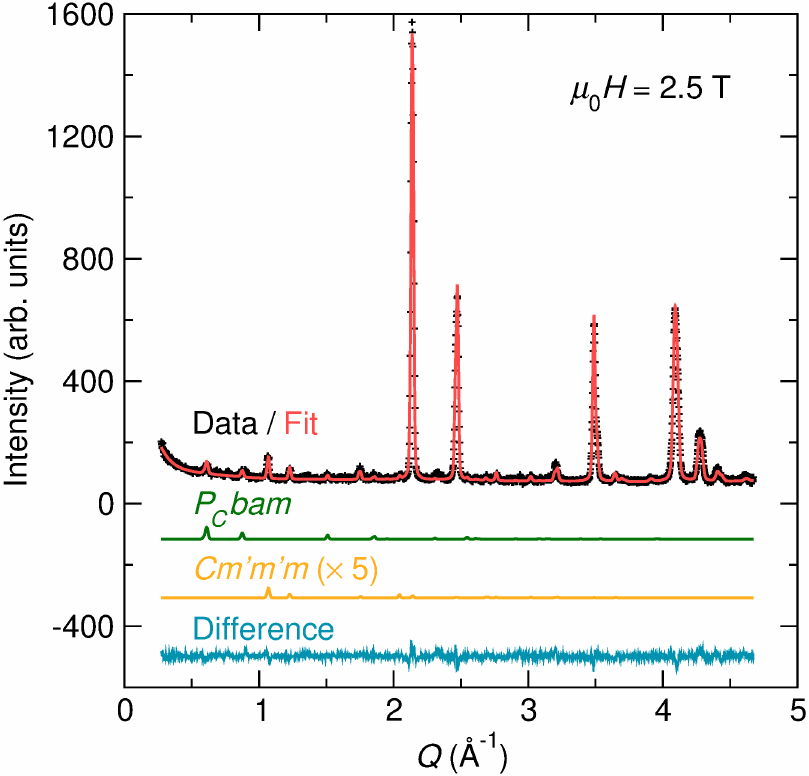}}
    \hspace{1.25cm}
    \subfloat{\includegraphics[width=0.4\columnwidth]{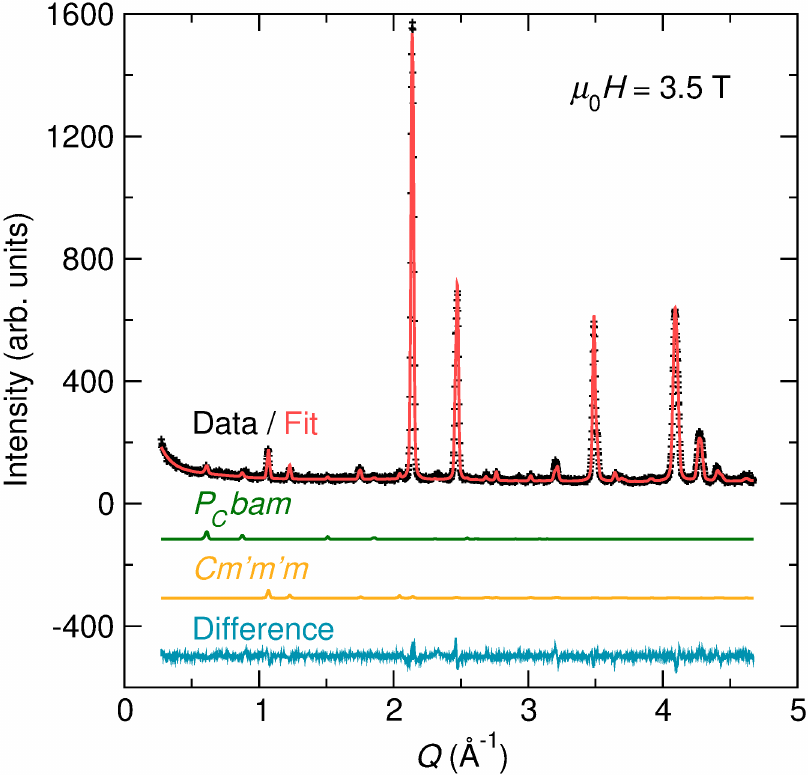}}
    \vspace{0.5cm}
    \subfloat{\includegraphics[width=0.4\columnwidth]{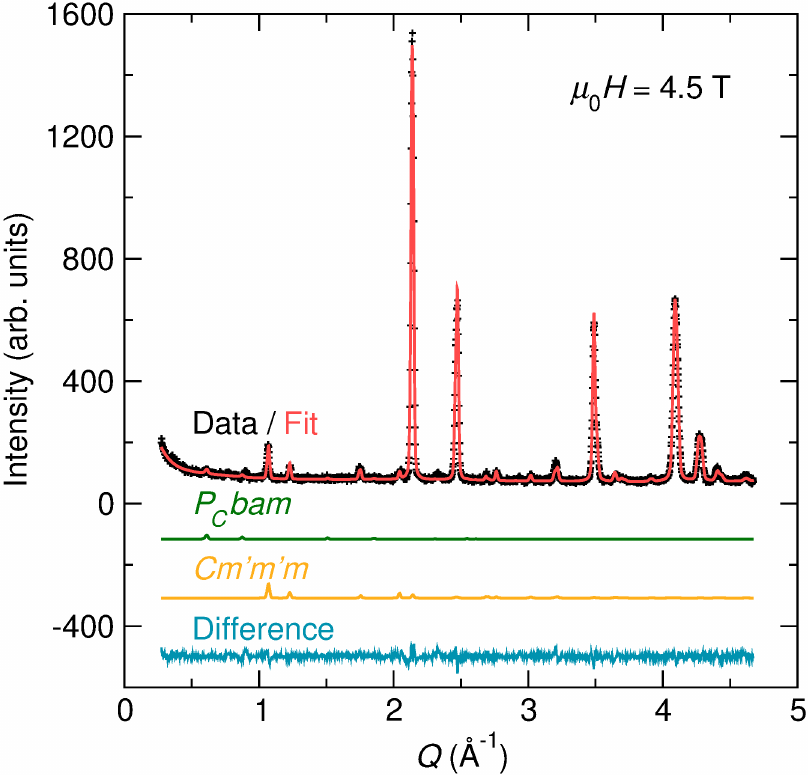}}
    \hspace{1.25cm}
    \subfloat{\includegraphics[width=0.4\columnwidth]{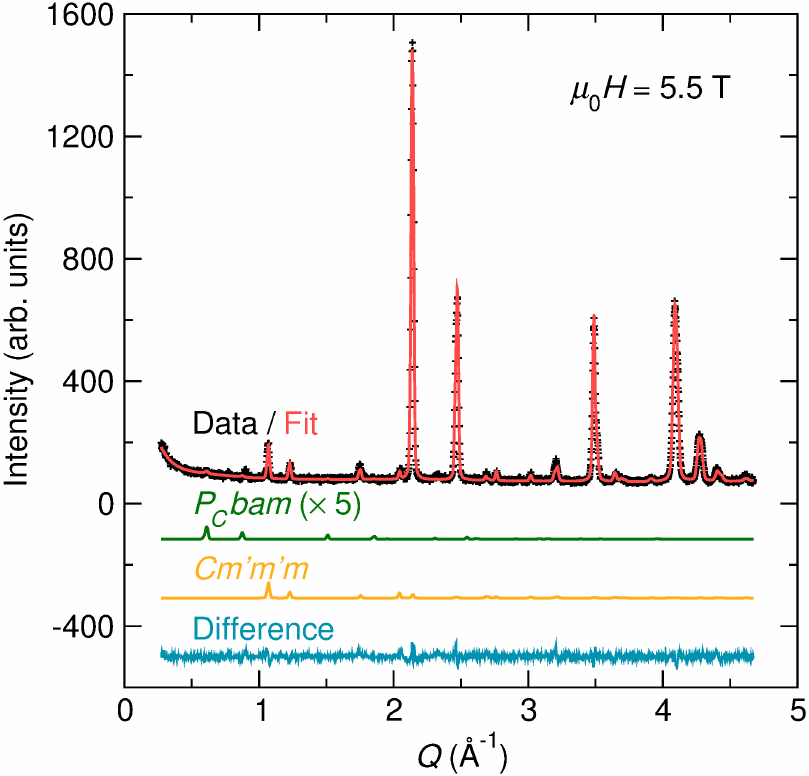}}
    \caption{Neutron powder diffraction refinements not in the main text are shown along with full pattern refinements of the 2.5 and 5.5~T data. At 0 and 1.5~T, we observe only antiferromagnetic reflections ($P_{C}bam$). At higher fields, the ferromagnetic phase co-exists with the antiferromagnetic one.}
    \label{fig:other_npd_fits}
\end{figure*}

\begin{figure*}
    \centering
    \subfloat{\includegraphics[width=0.475\columnwidth]{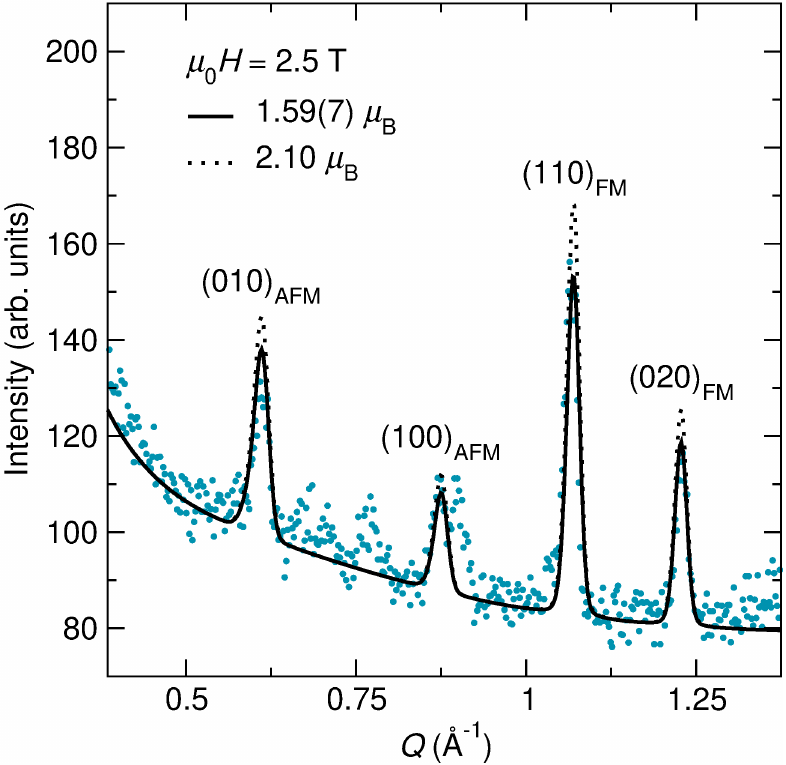}}
    \hspace{0.6cm}
    \subfloat{\includegraphics[width=0.475\columnwidth]{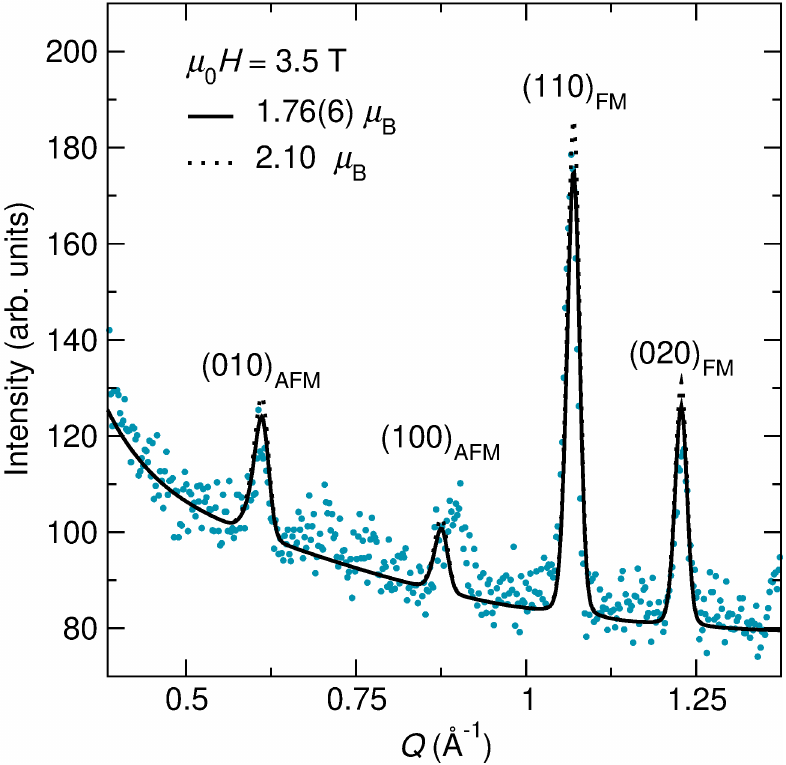}}
    \caption{Because of the intermediate-field, {\color{black}spin-reorientation} transition, the difference between refining the magnetic moment or constraining it to the zero-field value of 2.10~${\mu}_{\mathrm{B}}$ is most prominent for the highest intensity antiferromagnetic (``AFM") and ferromagnetic/nuclear (``FM") peaks in the 2.5 and 3.5~T data. For the 0, 0.8, 1.5, 4.5, and 5.5~T refinements, the difference is negligible. Unfit peaks are background/impurity peaks noted in the main text.}
    \label{fig:const_moment_changes}
\end{figure*}

\begin{figure}
    \centering
    \includegraphics[width=0.65\columnwidth]{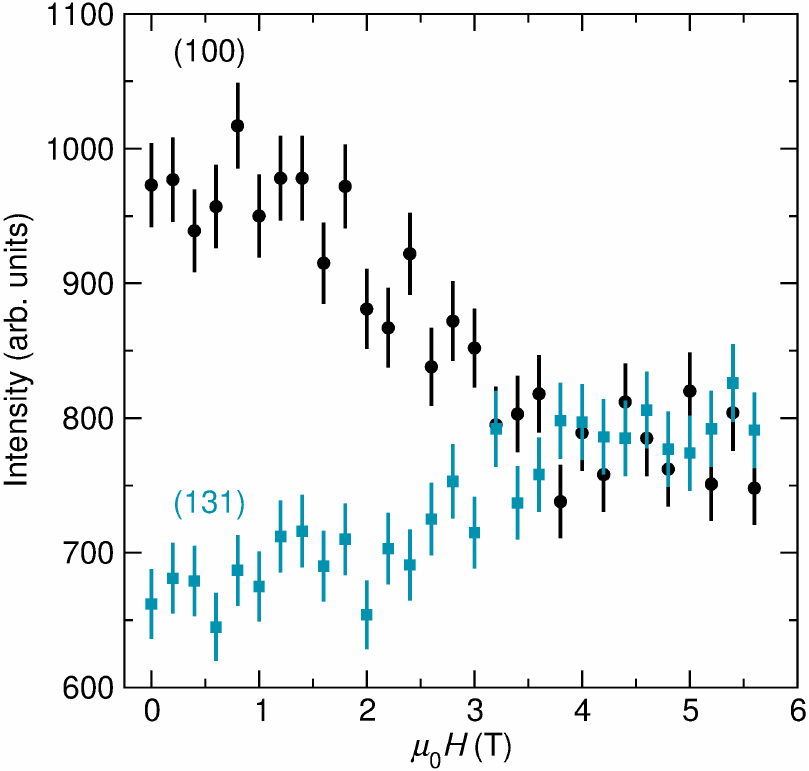}
    \caption{While collecting a magnetic field sweep of the prominent (010) antiferromagnetic reflection, other detectors captured information about the (100) antiferromagnetic reflection and the (131) nuclear/ferromagnetic reflection. Neither detector was centered on the (100) or (131) peak's maximum, but both captured the trend toward moments aligned between Co$^{2+}$ chains with increasing field. The convergence of both peaks to similar intensities is a coincidence of the higher background intensity around the (100) peak and the relatively low intensity of the (131) peak.}
    \label{fig:npd_xtra_sweeps}
\end{figure}

\clearpage
\section{Other symmetry options for the ferromagnetic phase}
\begin{figure}[h]
    \centering
    \includegraphics[width=0.65\columnwidth]{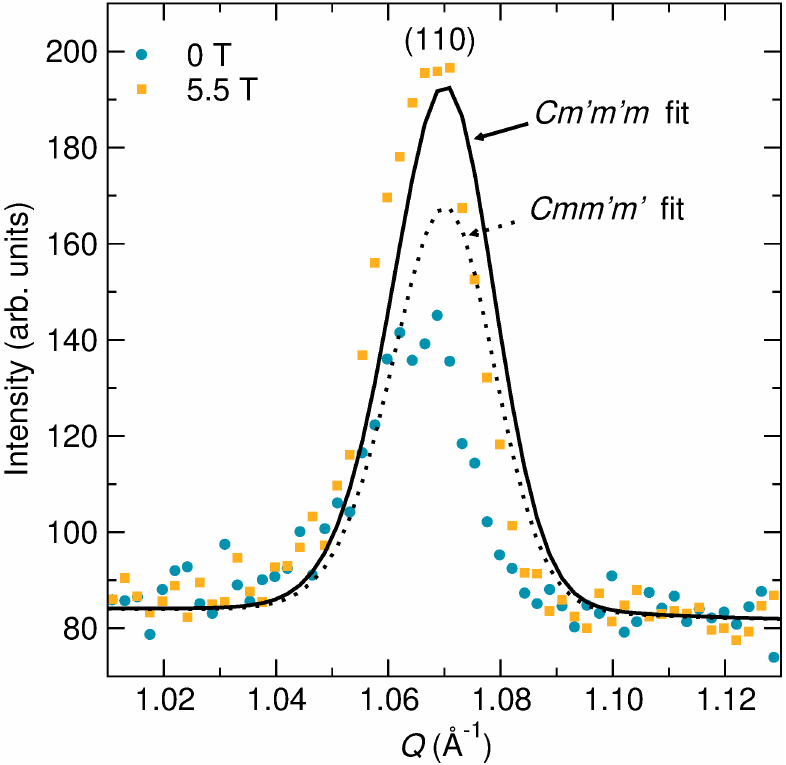}
    \caption{The magnetic intensity gained by the (110) peak with increasing field is better fit using a ferromagnetic phase with moments along the $c$ axis ($Cm'm'm$) than with moments along the $a$ axis ($Cmm'm'$).}
    \label{fig:110_change}
\end{figure}

{\color{black}In the main text refinements, we model the ferromagnetic phase that emerges in the polycrystalline sample with increasing field with the maximal magnetic space group $Cm'm'm$, where moments are restricted to the $c$-axis direction.} {\color{black}Several} factors oppose modeling the magnetic structure with the other maximal groups $Cmm'm'$ (moments along $a$ only) and $Cm'mm'$ (moments along $b$ only). First, the increasing intensity on the (020) reflection violates the $Cm'mm'$ systematic absence of (0\textit{k}0) reflections. Second, at higher fields, the significant increase in intensity on the (110) reflection (Fig.~\ref{fig:110_change}), though allowed for both the $Cm'm'm$ and $Cmm'm'$ symmetries, suggests a preferred moment alignment along $c$. Only the moment component within the (110) plane, and thus perpendicular to the corresponding scattering vector, will contribute to the magnetic intensity. For moments along $c$, this component equals the magnitude of the total moment for the $Cm'm'm$ symmetry structure but is less than the magnitude of the total moment for the $Cmm'm'$ symmetry structure. Consequently, fitting with $Cmm'm'$ does not account for the full (110) reflection intensity increase (Fig.~\ref{fig:110_change}). Third, a model with moments along $c$ fits nicely with the previously established zero-field structure and the proposed {\color{black}spin-reorientation} {\color{black}transition.} 

\clearpage
\section{High-field magnetic susceptibility and neutron data}
\begin{figure*}[h]
    \centering
    \subfloat{\includegraphics[width=0.45\columnwidth]{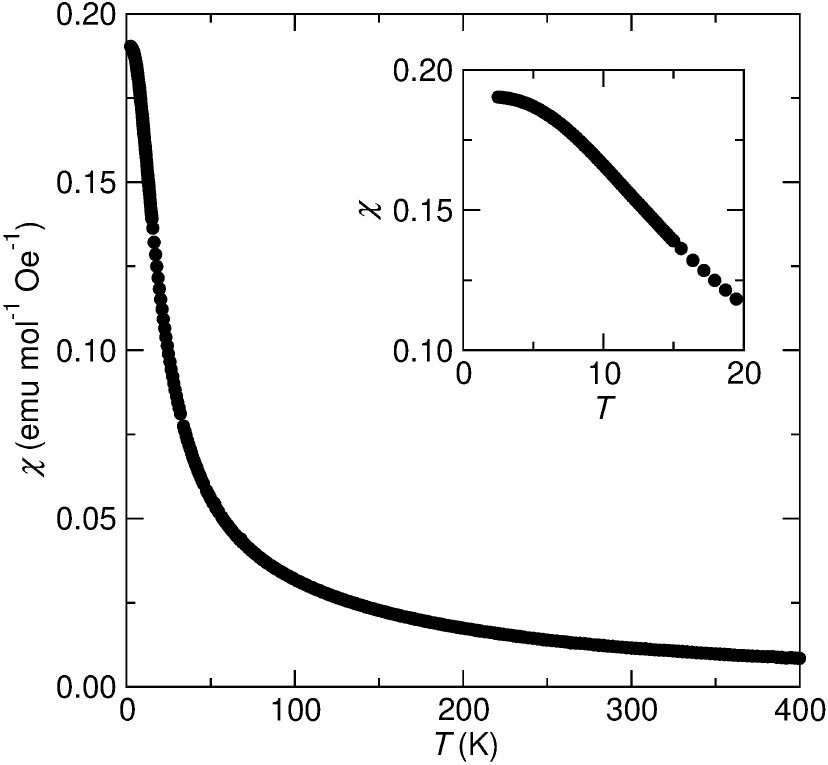}}
    \hspace{1.25cm}
    \subfloat{\includegraphics[width=0.436\columnwidth]{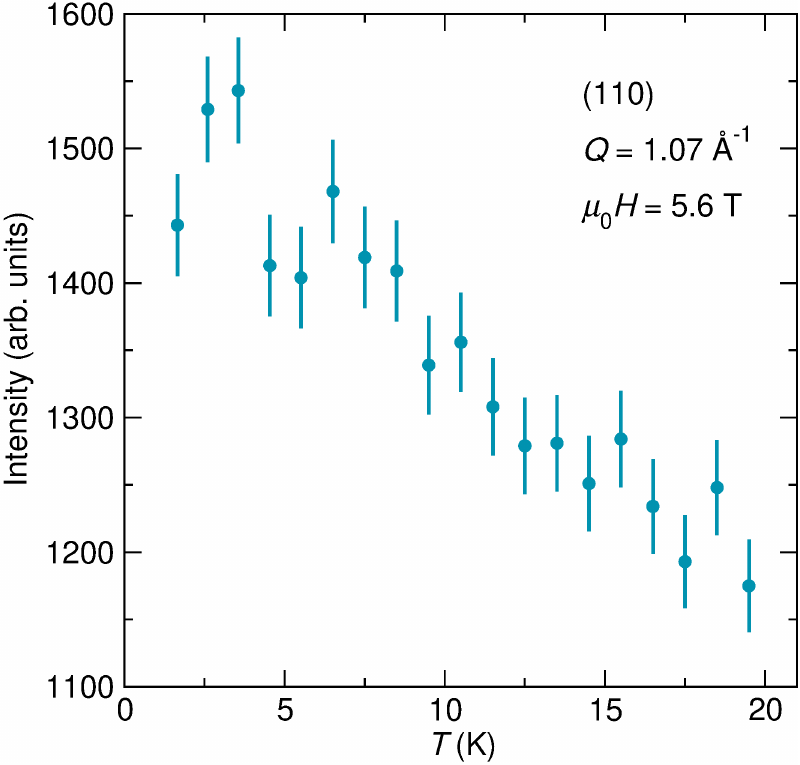}}
    \caption{(left) The zero-field cooled magnetic susceptibility of \LCC\ was measured at 70~kOe from 2.5 to 400~K in a Quantum Design Magnetic Property Measurement System (MPMS3), using the direct current measurement mode. (right) The neutron diffraction (110) reflection intensity is tracked with increasing temperature at 5.6~T.}
    \label{fig:chiZFC-7T_110-5p6T}
\end{figure*}

{\color{black}Fig.~\ref{fig:chiZFC-7T_110-5p6T} contains the temperature-dependent magnetic susceptibility of polycrystalline \LCC\ at 7~T and the temperature-dependent intensity of the ferromagnetic (110) reflection in neutron diffraction. The susceptibility levels off near the base temperature, with no sharp rise associated with purely ferromagnetic order and no cusp associated with long-range antiferromagnetic order. Similarly, the nuclear/ferromagnetic (110) reflection intensity drops nearly linearly with increasing temperature and does not show a critical point below 19.5~K. Both measurements indicate a competition between the applied magnetic field, grain orientations, ferromagnetic interactions, and antiferromagnetic interactions.}

\clearpage

\section{Additional IR data}
\begin{figure}[h]
    \centering
    \includegraphics[width=\columnwidth]{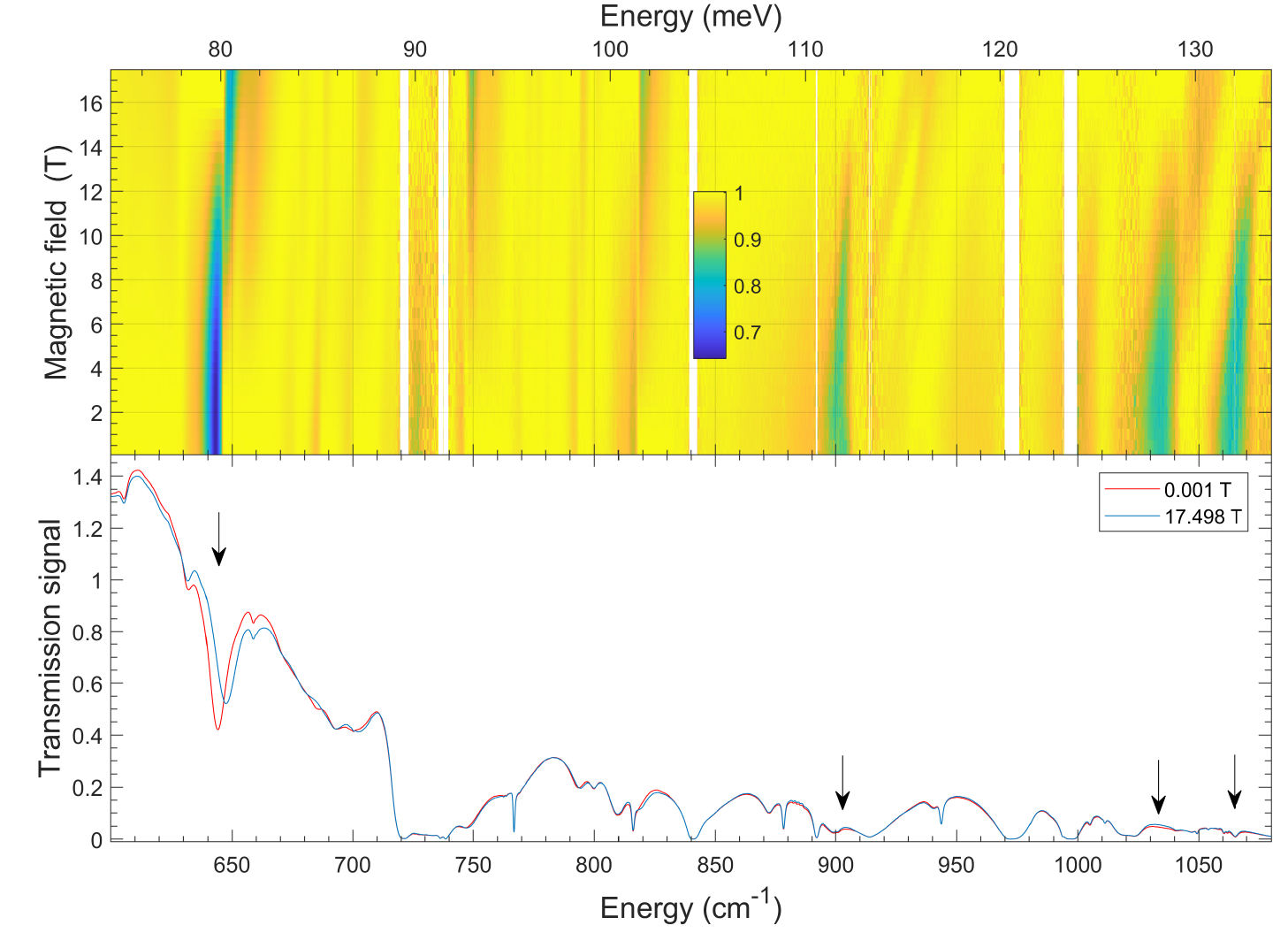}
    \caption{(top) Heatmap of the normalized transmittance, where the color represents relative changes in the (bottom) powder transmission signal induced by the magnetic field. The decrease in transmission at high frequencies due to spectrometer settings makes it difficult to distinguish between the 0 and 17.5~T spectra. Instead, normalizing all spectra to the reference spectrum highlights the relevant field-induced changes. The arrows indicate the zero-field positions of the most intense spectral features, which we attribute to transitions between Kramers doublets of the orbital $^{4}T_{\mathrm{1} g}$ state. }
    \label{fig:IRbroad}
\end{figure}

\clearpage
\section{Simulation of IR data with free-ion models}
\begin{figure}[h]
    \centering
    \includegraphics[width=\columnwidth]{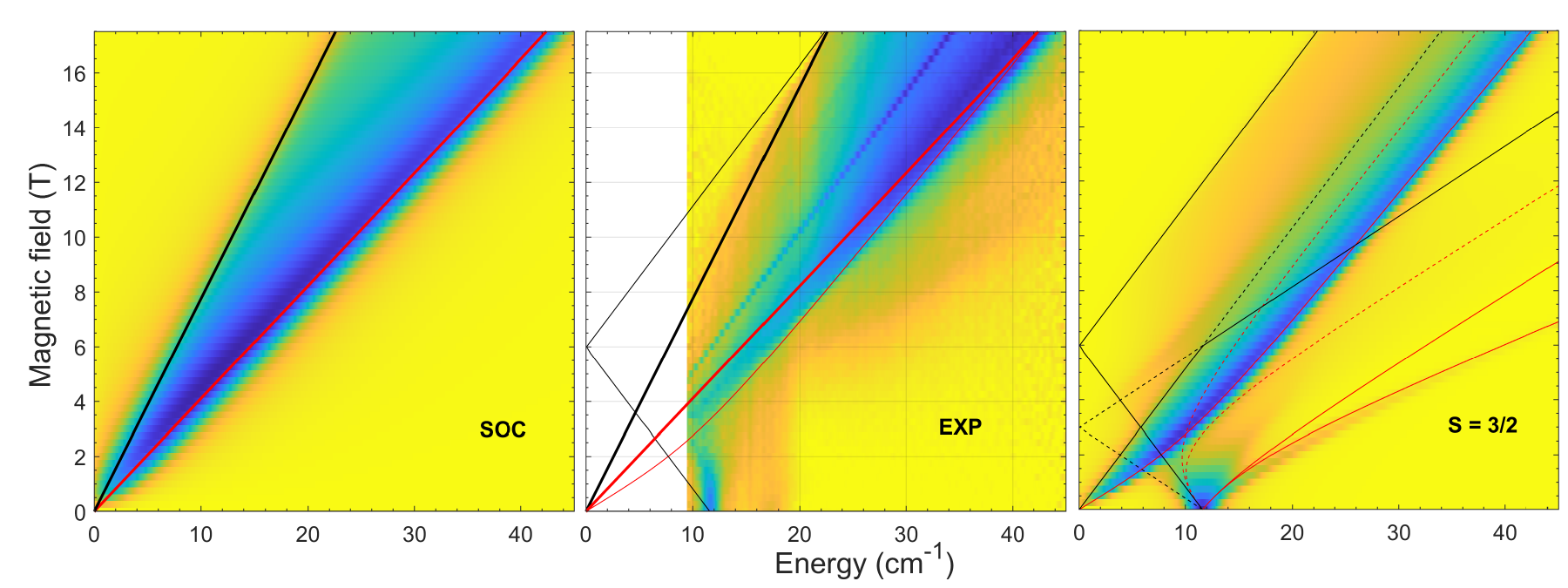}
    \caption{Comparison of the experimental magnetic resonance absorption heatmap (middle) with simulation results obtained using Hamiltonians for the free-ion approximation with spin-orbit coupling (left) and spin-$\frac{3}{2}$ (right) models for a powder sample using the \textsc{EasySpin} package \cite{stoll2006easyspin}. The parameters of the Hamiltonian are given in Table~\ref{tab:spectr}. The solid and dashed lines present the resonance excitations from the ground and first excited states respectively. The colored lines restrict the area of all possible magnetic resonances in polycrystalline samples, corresponding to the resonance dependence for magnetic field applied along $g_{\parallel}$ (black) and  $g_{\perp}$ (red). The broadening shown in the spin-orbit coupling case is the better match to the 17.5~T experimental data.}
    \label{fig:Sim}
\end{figure}
\vspace{3em}
\begin{table}[h!]
\small
\centering 
\caption{\label{tab:spectr} Spin Hamiltonian parameters used to simulate the magnetic resonance absorption spectra shown in Fig.~\ref{fig:Sim}.} 

\begin{tabular}{c c c c c}

\midrule
Model& $g_\perp$ &$g_\parallel$ & $[D,E]$ & $[\alpha,\lambda,\Delta/(\lambda\alpha)]$  \\
\midrule 

SOC   & 2   & 2    & ---     & 1.77, 157, 2 \\
$S~=~\frac{3}{2}$ & 4.5 & 4.16 & 5.8, 0  & --- \\
\end{tabular}
\end{table}

\clearpage

\section{Evaluation of the exchange interaction anisotropy for an effective Heisenberg Hamiltonian model}
Following Abragam and Pryce \cite{abragam1951theory}, the wavefunction of the ground state doublet for the spin-orbit Hamiltonian (Eq.1 in main text) is expressed with orbital-spin wavefunctions $\ket{L,S}$ as

\begin{equation}
    \label{eq:wave}
    \ket{\pm\frac{1}{2}} = a \ket{\mp1,\pm\frac{3}{2}} + b\ket{0,\pm\frac{1}{2}}+c\ket{\pm1,\mp\frac{1}{2}}
\end{equation}
where $a$, $b$, and $c $ are coefficients dependent on $\Delta/(\lambda\alpha)$. If we use the spectroscopically estimated $\Delta$, $\alpha$, and $\lambda$ parameters (listed in the main text and in Table~\ref{tab:spectr}), we find 

\begin{equation} \label{eq:waveabc}
 \begin{split}
    a= -0.4931; b= 0.7844; c=-0.3763
\end{split}
\end{equation}

When the ground state is well separated from the first excited state, the total spin $S_\mathrm{tot}=\frac{3}{2}$ within the ground state doublet can be described by a $S=\frac{1}{2}$ function

\begin{equation} \label{eq:effSpin0}
    S_\mathrm{tot}^{x,y}=qS^{x,y}, S_\mathrm{tot}^z=pS^z
\end{equation}
 where $q$ and $p$ are coefficients defined as in Ref.~\cite{oguchi1965theory}.
\begin{equation} \label{eq:coeff}
\begin{split}  
    p=3a^2+b^2-c^2= 1.2032\\
    q=2b^2+2\sqrt{3} ac=1.8733
    \end{split}
    \end{equation}
    
Hence, the isotropic superexchange mechanism for neighboring total spins will be mapped to an effective $S=\frac{1}{2}$ Hamiltonian with an anisotropic exchange interaction  

\begin{equation} \label{eq:effSpin1}
    H_{XXZ}=J \sum_{i=1}^{N} (S_i^xS_{i+1}^x+S_i^yS_{i+1}^y +\beta S_i^zS_{i+1}^z)
\end{equation}
where 

\begin{equation} \label{eq:beta}
    \beta=\frac{p^2}{q^2}=0.4125
\end{equation}

\begin{figure}
    \centering
    \includegraphics[width=\columnwidth]{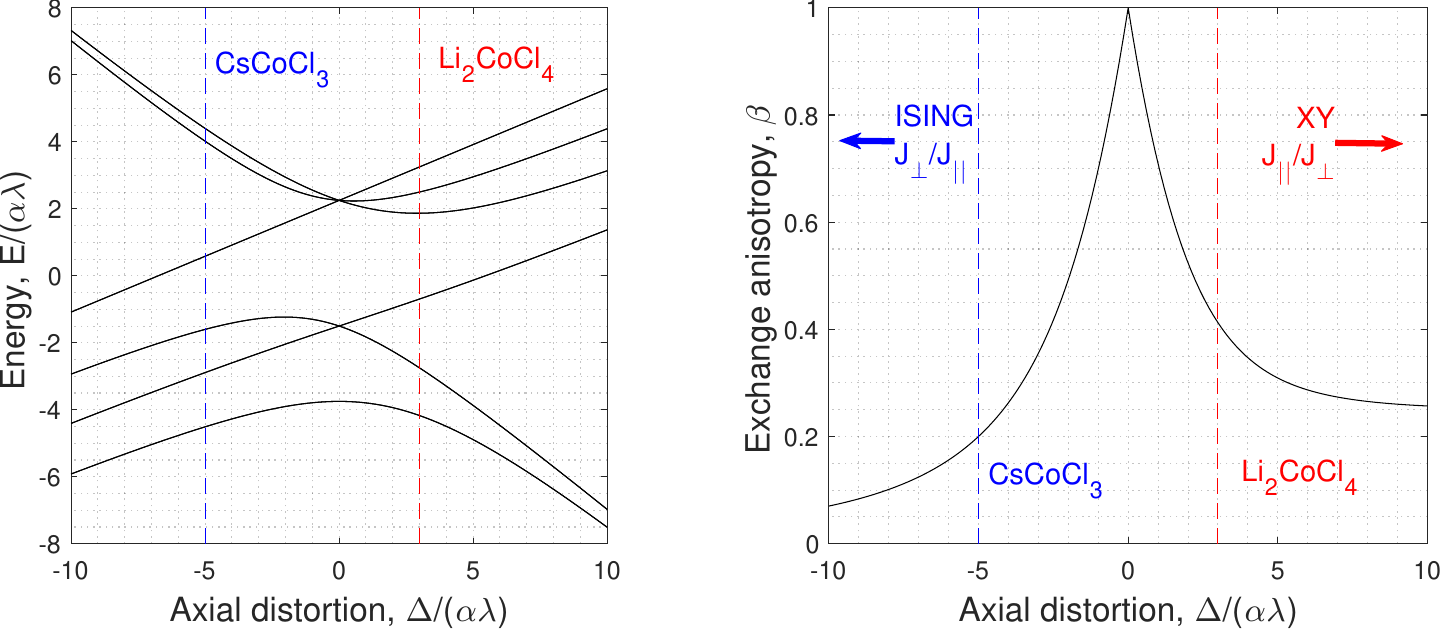}
    \caption{(left) Energy level splitting of the orbital triplet ground state of the Co$^{2+}$ ion ($^4T_{1g}$) due to axial distortion. (right) Ratio of the parallel and perpendicular components of the nearest-neighbor exchange interaction for a $S$=$\frac{1}{2}$ single-chain effective Hamiltonian. Dashed lines show the estimated distortion values for \LCC\ (this work) and CsCoCl$_3$ \cite{achiwa1969linear}.}
    \label{fig:anisotropy}
\end{figure}

This approach for estimating the exchange anisotropy has previously been applied for the CsCoCl$_3$ quasi-1D Ising system \cite{achiwa1969linear,shiba2003exchange}. A comparison is shown in Fig.~\ref{fig:anisotropy}. Note that the energy and axial distortion of our calculations were multiplied by a factor of 1.5 to make our results compatible with Fig.~1 and Fig.~3 in Ref.~\cite{shiba2003exchange}. The factor of 1.5 is a coefficient in front of spin-orbit coupling term in the Hamiltonian used in Refs.~\cite{achiwa1969linear, shiba2003exchange}.

\clearpage

\bibliography{Li2CoCl4.bib}